\documentclass[%
 reprint,
superscriptaddress,
 amsmath,amssymb,
 aps,
prl,
]{revtex4-2}

\usepackage{graphicx}% 
\usepackage{dcolumn}% 
\usepackage{multirow}
\usepackage{amssymb}
\usepackage{comment}
\usepackage{MnSymbol,wasysym}
 \usepackage{setspace}

\usepackage[dvipsnames,table,xcdraw]{xcolor}

\usepackage{hyperref}
\hypersetup{
    colorlinks=true,
    linkcolor=Blue,
    filecolor=Blue,      
    urlcolor=Blue,
    citecolor=blue,
}

\definecolor{Gray}{gray}{0.85}
\newcolumntype{g}{{\columncolor{Gray}}r}
\newcolumntype{w}{{\columncolor{white}}r}

\def\bQ{{\mathbf{Q}}}
\def\br{{\mathbf{r}}}
\renewcommand\vec[1]{\ensuremath\boldsymbol{#1}} % bold font for vectors

\DeclareUnicodeCharacter{2212}{-}

\usepackage[nopar]{lipsum}

\begin{document}

%%%%%%%%%%%%%%%%%%%%%%%%%%%%%%%%%%%%%%%%%
\title{
Intertwining bulk and surface: 
the case of UTe$_2$
}

\author{Andr\'{a}s L. Szab\'{o}}
\affiliation{Institute for Theoretical Physics, ETH Zurich, 8093 Zurich, Switzerland}

\author{Aline Ramires }
\affiliation{Laboratory for Theoretical and Computational Physics, PSI Center for Scientific Computing, Theory, and Data,  Paul Scherrer Institute, 5232 Villigen PSI, Switzerland.}
\thanks{Current address: Institute of Solid State Physics, TU Wien, 1040 Vienna, Austria. Email: aline.ramires@tuwien.ac.at}

%\curraddr{Institut f\"{u}r Festk\"{o}rperphysik, TU Wien, 1040 Vienna, Austria}

\date{\today}

\begin{abstract}
UTe$_2$ has been the focus of numerous experimental and theoretical studies in recent years, as it is recognized as an odd-parity bulk superconductor. 
Its surface has also been probed, revealing charge density wave (CDW), pair density wave (PDW), and time-reversal symmetry breaking (TRSB). 
In this work, we propose that the interplay between the order parameters observed on the surface and in the bulk of UTe$_2$ may be crucial in explaining some of the unusual features detected by surface probes in this material. 
Through a phenomenological analysis, we can account for three distinctive experimental signatures observed on the surface of UTe$_2$: 
i) the apparent suppression of CDW order at the upper critical field of the bulk superconducting state; 
ii) the magnetic field-induced imbalance of the Fourier peaks associated with the CDW; 
iii) the onset of TRSB at the bulk superconducting critical temperature and its field-trainability. 
Furthermore, we propose specific experimental checks to validate our conjecture, which we believe could be promptly achieved.

\end{abstract}

\maketitle 

\emph{Introduction.}
Modulated orders in the charge, spin, and superconducting sectors are ubiquitous in strongly correlated systems and are associated with a plethora of non-trivial phenomena such as intertwined and vestigial orders \cite{Fradkin2015,Fernandes2019}. Focusing on the superconducting state, modulated orders are known as pair-density waves (PDWs)~\cite{Agterberg2020}. When PDWs are the primary orders, they can induce multiple secondary orders with a remarkable interplay of topological defects~\cite{Agterberg2008, Berg2009NP, Berg2009NJP}. PDWs can also emerge as secondary or induced orders, in which case their symmetries can reveal signatures of the underlying primary orders that could otherwise be hard to detect.

PDWs have been primarily characterized by scanning tunneling microscopy (STM), an intrinsically surface-sensitive technique.
STM experiments originally reported PDWs in high-temperature cuprate superconductors~\cite{Hamidian2016, Ruan2018, Edkins2019}. 
Most recently, NbSe$_2$, a transition-metal dichalcogenide~\cite{Liu2021}, and CsV$_3$Sb$_5$, a hexagonal kagome material~\cite{Chen2021} have been reported to host exotic PDWs.
Most notable is the case of UTe$_2$, for which STM experiments have reported a PDW~\cite{Gu2023} and an unusual magnetic-field dependent CDW that seems to be suppressed at the bulk superconducting upper critical field \cite{Aishwarya2023, LaFleur2024}. 
While superconductivity is a well established bulk phase in this material, x-ray and thermodynamic measurements failed to identify signatures of CDW in the bulk~\cite{Kengle2024natcom,Theuss2024,Kengle2024}.

\begin{figure}[t]
    \centering
    \includegraphics[width=0.45\textwidth]{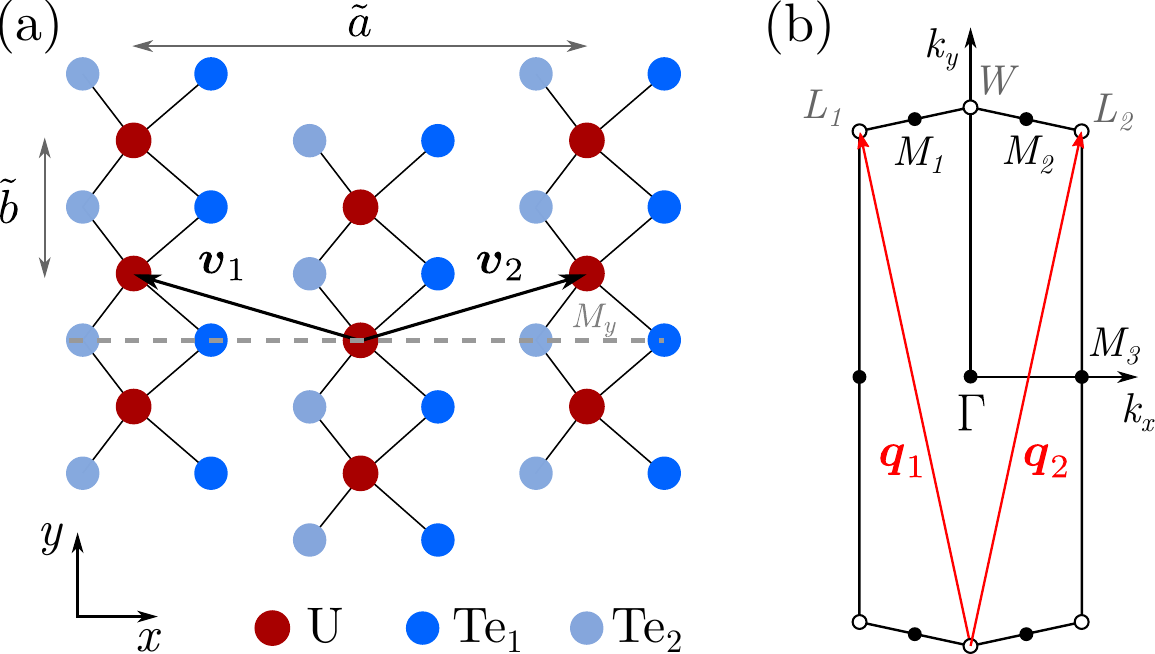}
    \caption{a) Easy-cleave surface of UTe$_2$ with lattice constants $\tilde{a}$ and $\tilde{b}$ along the surface $x$- and $y$-directions, respectively. The basis vectors $\vec{v}_1$ and $\vec{v}_2$ are indicated by the black arrows.
    The dashed gray line indicates the $M_y$ mirror plane.
    b) Surface BZ highlighting high-symmetry points (filled circles) and other relevant points (open circles). The reciprocal lattice vectors $\boldsymbol{q}_1$ and $\boldsymbol{q}_2$ are indicated by the red arrows.}
    \label{fig:SurfaceSymmetry}
\end{figure}

A growing body of experimental evidence indicates that UTe$_2$ is an odd-parity spin-triplet bulk superconductor \cite{Ran2019, Aoki2019, Nakamine2019}.
In addition, signatures of time-reversal symmetry  breaking (TRSB) \cite{Sundar2019, Azari2023, Hayes2021, Wei2022}, the chiral nature of STM spectra at step edges \cite{Jiao2020}, and the anisotropy of low-energy excitations \cite{Ishihara2023} are features that generally require a multicomponent order parameter.
However, the low, orthorhombic symmetry of UTe$_2$ does not host higher-dimensional irreducible representations (irreps) supporting multicomponent order parameters.
In this light, theoretical works have proposed a variety of accidental degeneracies or coexistence of symmetry-distinct homogeneous superconducting order parameters to account for the unusual phenomenology of UTe$_2$ from the bulk perspective \cite{Aoki2022, Theuss2024NaturePhysics}.
Most recently, the connection between bulk and surface has been explored by quasi-particle interference~\cite{Wang2025,Gu2025} and theoretical calculations~\cite{Crépieux2025,Christiansen2025}, suggesting a $B_{3u}$ order parameter.

Here, we take an alternative approach to address the phenomenology of this material: we show that the intertwining of bulk homogeneous single-component superconductivity with surface CDW gives rise to a multicomponent PDW on the surface that can account for much of the unusual observations on the surface of UTe$_2$.
Using a phenomenological Landau theory, we address the suppression of CDW order as a function of magnetic field; propose a scenario for TRSB superconducting state based on the induced surface PDW; and unravel the  origin of the increased imbalance of the Fourier peaks capturing modulated orders in STM experiments in the presence of an external magnetic field.

\emph{Symmetry considerations.}
UTe$_2$ is a three-dimensional heavy-fermion system crystallizing in space group $I/mmm$ ($\# 71$), with a body-centered orthorhombic structure \cite{Hutanu2020}. 
STM experiments have probed the easy-cleave $(0\bar{1}1)$ surface, characterized by wallpaper group $cm$, which includes a mirror operation $M_y$, with normal along the $y$ direction, 
see Fig. \ref{fig:SurfaceSymmetry}(a).
The primitive lattice vectors $\vec{v}_{1,2} = (\mp \tilde{a}/2, \tilde{b}/2)$, where $\tilde{a}$ and $\tilde{b}$ are the surface lattice parameters, span a 
centered rectangular
%squeezed triangular 
lattice, resulting in an elongated hexagonal BZ, with three high-symmetry points,  $M_{1,2,3}$, see Fig.~\ref{fig:SurfaceSymmetry} (b). 
In addition to the high-symmetry points, we also label the vertices of the BZ as $W$ and $L_{1,2}$ following the notation in Aishwarya et al. \cite{Aishwarya2023, Setyawan2010}.

The CDW order observed at the UTe$_2$ surface has been associated with incommensurate wave vectors $\vec{Q}_{i=0,1,2}$ near the $W$ and $L_{1,2}$ points of the BZ, respectively,
with some of these wave vectors are related by reciprocal lattice vectors (RLVs) $\vec{q}_{1,2}=2\pi(\mp 1/\tilde{a},1/\tilde{b})$, as $\pm \vec{Q}_{1,2}=\mp \vec{Q}_0 \pm \vec{q}_{1,2}$.
The observed modulated orders, therefore, correspond to multiple ``harmonics'' of a two-component order parameter, related by RLVs.
For the discussions associated with global symmetries, it suffices to focus on only two independent wave vectors, which we choose to be $\vec{Q}_0$ and $-\vec{Q}_0= M_y\vec{Q}_0$, associated with the two-component CDW order parameter $\mathbf{\Delta}_0=(\Delta_{+Q_0}, \Delta_{-Q_0})$, leading to a spatial charge distribution following $\Delta(\br) = \Delta_{+Q_0} e^{i \mathbf{Q}_0\cdot \br} + \Delta_{-Q_0} e^{-i \mathbf{Q}_0\cdot \br}$.

\emph{Landau theory.}
We start the phenomenological description of the surface with the CDW order, which has the highest onset temperature with $T_{{\rm c},\Delta}\approx 12$K~\cite{LaFleur2024}.
Choosing $\pm\vec{Q}_0$ as the representative harmonics, in addition to the CDW order parameter $\mathbf{\Delta}_0$, a modulated current order parameter $\mathbf{\Delta}_0'=(\Delta'_{+Q_0}, \Delta'_{-Q_0})$ is induced in the presence of a magnetic field.
Here $\mathbf{\Delta}_0$ and $\mathbf{\Delta}_0'$ are, respectively, even and odd under time-reversal symmetry (TRS), such that $\Delta_{+Q_0}^*=\Delta_{-Q_0}$ and $\Delta_{+Q_0}^{\prime *}=-\Delta'_{-Q_0}$.
In addition to being odd under TRS, magnetic field contained in the mirror plane $\vec{B}_{\parallel M}=(B_x, 0, B_z)$ is odd, while field perpendicular to the mirror $\vec{B}_{\perp M}=(0, B_y, 0)$ is even under $M_y$.
For the sake of simplicity, we set $B_x=0$ from here on.
%, see SM.
The Landau free energy in the charge-current sector up to fourth order reads
\begin{align}
    &F_{\Delta}=\frac{\alpha}{2} |\mathbf{\Delta}_0|^2 + \frac{\beta}{4} |\mathbf{\Delta}_0|^4 + 
      \frac{\alpha'}{2} |\mathbf{\Delta}'_0|^2 + \frac{\beta'}{4} |\mathbf{\Delta}'_0|^4  \label{eq:FreeEnergyCDW}\\
      &+\lambda B_y \mathbf{\Delta}_0^\dagger \mathbf{\Delta}_0' + \eta B_z \mathbf{\Delta}_0^\dagger \sigma_3 \mathbf{\Delta}_0' + B_z^2\left(\frac{\zeta}{2} |\mathbf{\Delta}_0|^2 + \frac{\zeta'}{2} |\mathbf{\Delta}_0'|^2\right), \nonumber
\end{align}
where $\alpha^{(\prime)}=a^{(\prime)}(T-T_{{\rm c},\Delta^{(\prime)}})$, with $T_{{\rm c},\Delta^{(')}}$ the critical temperature of the pure charge (current) density wave. $\sigma_i$ are Pauli matrices, and we assume the phenomenological coefficients $a^{(\prime)},\beta^{(\prime)}, \zeta^{(\prime)}, \lambda, \eta>0$. 
We also assume there is no spontaneous transition to a phase with $|\mathbf{\Delta}_0'|\neq 0$ in absence of magnetic field, such that $\alpha'>0$.

\begin{figure}
    \centering
    \includegraphics[width=0.98\linewidth]{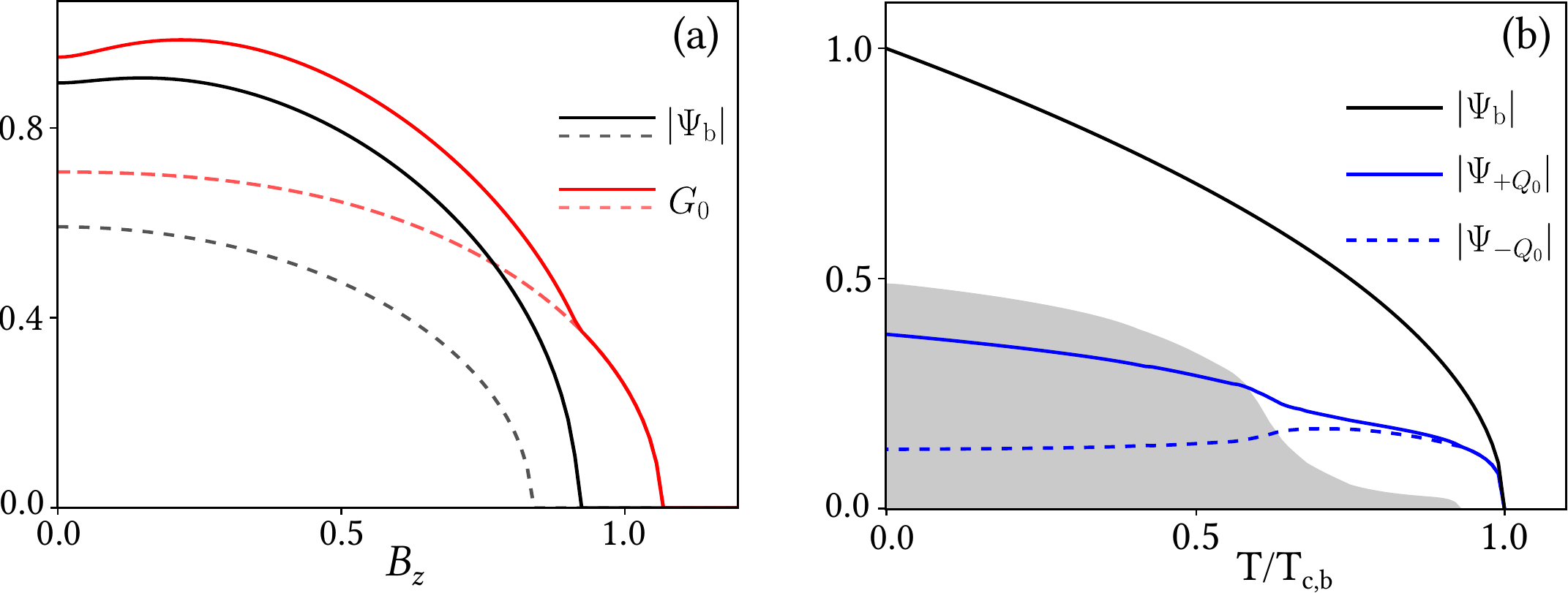}
    \caption{(a) Evolution of bulk superconducting order, $|\Psi_{\rm b}|$, and combined CDW and field-induced current order ($G_0$) as a function of $B_z$. Dashed and solid lines correspond to $\gamma=\gamma'=0$ (decoupled) and $\gamma/\beta=\gamma'/\beta=-1$ (attractively coupled), respectively.
    We used $\alpha'/|\alpha|=\beta'/\beta=\beta_{\rm b}/\beta=\zeta/\beta=\zeta'/\beta=\zeta_{\rm b}/\beta=1$, and $-\alpha_{\rm b}/|\alpha|=\eta/\sqrt{|\alpha|\beta}=0.5$.
    (b) TRSB via the imbalance of PDW components, $|\Psi_{\pm Q_0}|$. We quantify TRSB by $(|\Psi_{+Q_0}|-|\Psi_{-Q_0}|)/(|\Psi_{+Q_0}|+|\Psi_{-Q_0}|)$, shown as the shaded area. We used  $|\vec{\Delta}_0|/\sqrt{|\alpha|/\beta}=-\alpha_{\rm b}/|\alpha|=\beta_{\rm b}/\beta=1$, $\tilde{\alpha}/|\alpha|=\kappa/\sqrt{|\alpha| \beta}=0.1$, $\tilde{\beta}_1/\beta=1.3$, $\tilde{\beta}_2/\beta=-1$.
    }
    \label{fig:Intertwining}
\end{figure}

In the superconducting sector, we assume a single-component homogeneous bulk order parameter $\Psi_{\rm b}=|\Psi_{\rm b}|e^{i \phi_{\rm b}}$, which may be even or odd under $M_y$, and write its free energy as
\begin{align}
    &F_\Psi=
    \alpha_{\rm b} |\Psi_{\rm b}|^2 + \beta_{\rm b} |\Psi_{\rm b}|^4 + \zeta_{\rm b} B^2  |\Psi_{\rm b}|^2, 
    \label{eq:FreeEnergySC}
\end{align}
where  $\beta_{\rm b}>0$, $\alpha_{\rm b}=a_{\rm b}(T-T_{{\rm c},\Psi})$, with $a_{\rm b}>0$.
Here $T_{\rm c,\Psi}\approx1.5$K is the bulk superconducting critical temperature, about one order of magnitude smaller than $T_{{\rm c},\Delta}$, and $\zeta_{\rm b}>0$ captures the general suppression of the bulk superconducting state in the presence of an external magentic field.
The free energy also includes terms that couple both sectors discussed above, which read
\begin{align}
    &F_{\Delta \Psi}=
    \tilde{\alpha} |\mathbf{\Psi}_{0}|^2 + 
    \tilde{\beta}_1 |\mathbf{\Psi}_0|^4+ \tilde{\beta}_2|\mathbf{\Psi}_0^\dagger \sigma_3 \mathbf{\Psi}_0|^2 + \rho B_z\mathbf{\Psi}_0^\dagger \sigma_3 \mathbf{\Psi}_0
    \nonumber \\
    &+\frac{|\Psi_{\rm b}|^2}{2} \Big( \gamma |\mathbf{\Delta}_0|^2 + \gamma' |\mathbf{\Delta}'_0|^{2}  \Big)
    +
    \kappa \Big( \Psi^*_{\rm b} \mathbf{\Delta}_{0}^\dagger \sigma_{i} \mathbf{\Psi}_{0} \pm {\rm c.c.}\Big )  .
    \label{eq:FreeEnergyCoupled}
\end{align}
Here the $\gamma^{(\prime)}$ terms are simple biquadratic couplings of the CDW (current) and bulk superconducting orders, whereas the $\kappa$ term describes the induction of a two-component PDW order parameter $\mathbf{\Psi}_0 =(\Psi_{+Q_0},\Psi_{-Q_0})$ with $\tilde{\alpha}>0$, as we assume the PDW is not a stable phase in isolation.
The $\rho$ term captures the only linear coupling to magnetic field, and we keep terms up to fourth order in the PDW order parameters with $\tilde{\beta}_1>0$ and $\tilde{\beta}_1+\tilde{\beta}_2>0$ for stability.
The $\pm$ signs in the last line correspond to $\Psi_{\rm b}$ being even (i=0) or odd (i=3) under $M_y$.

\emph{Magnetic-field dependent CDW.} The Fourier peaks associated with the CDW at the surface of UTe$_2$ were reported to be rapidly suppressed under magnetic fields close to $H_{c2}$, the bulk superconducting upper critical field \cite{Aishwarya2023}.
In the light of the free energy above, we argue that the simplest explanation for this observation is that $\mathbf{\Delta}_0$ and $\Psi_{\rm b}$ are strongly and \emph{attractively} coupled, corresponding to a large negative $\gamma$ coefficient.
In this scenario, the CDW survives up to magnetic fields larger than the bulk upper critical field, but the onset of bulk superconductivity leads to a boost of $|\mathbf{\Delta}_0|$. Conversely, this feature can be interpreted as an apparent suppression of $|\mathbf{\Delta}_0|$ at $H_{c2}$, which we explicitly show by numerically minimizing the free energy, see Fig.~\ref{fig:Intertwining}(a).
Here we introduce the gap magnitude $G_i^2=\sum_{\tau=\pm}(|\Delta_{\tau Q_i}|^2+|\Delta'_{\tau Q_i}|^2)$ resulting from the presence of modulated charge and current orders with ordering vector $\pm \vec{Q}_i$.
Across this work, for numerical minimization we measure the free energy in units of $\alpha^2/\beta$, all order parameters and external fields in units of $\sqrt{|\alpha|/\beta}$, and use the dimensionless forms of phenomenological coefficients (specific values are given in the corresponding figure caption).

\emph{TRS breaking superconductivity.} In contrast to the single-component nature of the bulk superconducting order parameter, the intrinsic two-component nature of the PDW oder parameter $\mathbf{\Psi}_0$ could naturally lead to TRSB in the superconducting state.  
The spatial dependence of the PDW order parameter can be written as $\Psi_{\rm PDW} (\br) = \Psi_{+Q_0} e^{i \bQ_0\cdot \br } + \Psi_{-Q_0} e^{-i \bQ_0 \cdot \br}$.
Parametrizing the two components in terms of an overall magnitude $|\Psi_0|$, an angle $\theta$ controlling their relative magnitude, and angles $\phi_{\pm}$ capturing their phases as $\Psi_{+Q_0} = |\Psi_0| \cos\theta e^{i\phi_+}$ and $\Psi_{-Q_0} = |\Psi_0| \sin\theta e^{i\phi_-}$ ($0<\theta<\frac{\pi}{2}$), we can write explicitly:
\begin{align}
    \Psi_{\rm PDW} (\br) 
    &= |\Psi_0|e^{i\frac{(\phi_+ + \phi_-)}{2}}\left[\cos\theta e^{i \bQ_0\cdot\left( \br + \frac{(\phi_+ - \phi_-)}{2 |Q_0|} \hat{\bQ}_0 \right)} \right.
    \\ \nonumber
    &+ 
    \left.\sin\theta e^{-i \bQ_0 \cdot \left(\br + \frac{(\phi_+ - \phi_-)}{2 |Q_0|} \hat{\bQ}_0 \right)}\right],
\end{align}
where $\hat{\bQ}_0 = \bQ_0/|\bQ_0|$. 
Note that the relative phase between the two components, $(\phi_+-\phi_-)/2$, can be eliminated by a suitable choice of origin, as this corresponds to a shift of the PDW pattern in direct space. 
The average phase $(\phi_++\phi_-)/2$ corresponds to an overall phase of the PDW order parameter, which can be eliminated by a proper choice of gauge in the absence of other superconducting phases.  Note that, in contrast to the usual notion of TRSB in a two-component superconducting order parameter emerging from the relative complex phase between them \cite{Sigrist1991, Kallin2016, Ramires2022}, TRSB in the PDW manifests as an imbalance between the two components, namely, $\theta \neq \pi/4$.

Minimizing the total free energy $F = F_\Delta + F_\Psi + F_{\Delta\Psi}$ in the presence of both CDW and bulk superconductivity, we find an induced PDW with $|\mathbf{\Psi}_0| \approx \kappa |\Psi_b||\mathbf{\Delta}_0|/\tilde{\alpha}$. 
%To minimize w.r.t. $\theta$, we first focus on the $\kappa$ term only.
From here on, without loss of generality, we assume the choice of origin at a maximum of the CDW, such that $\Delta_{+Q_0} = \Delta_{-Q_0}^* = |\mathbf{\Delta}_0|/\sqrt{2}$ and gauge out the phase of the bulk superconducting state, such that $\Psi_{\rm b} = |\Psi_{\rm b}|$. 
Note that the $\kappa$ term in Eq.~\ref{eq:FreeEnergyCoupled} is extremized for $\theta=\pi/4$, inducing a PDW with two components of same magnitude.
This corresponds to a TRS-preserving PDW, as  expected, since the bulk SC state and CDW do not break TRS.
For $\Psi_{\rm b}$ even under $M_y$, the free energy is minimized for $\phi_\pm = \pi (0)$ for $\kappa>0$ ($\kappa<0$). Note that the difference $\phi_+-\phi_-$ in both cases is zero, implying that the amplitudes of the PDW and CDW are not shifted w.r.t. each other, as experimentally observed in STM experiments \cite{Gu2023}. 
Note also that the average phase for $\kappa>0$ ($\kappa<0$) indicates that the PDW order parameter has a $\pi$ (zero) phase shift with respect to the bulk superconducting state. 
For mirror-odd $\Psi_{\rm b}$, minimizing the free energy yields zero average phase and a $\pm \pi/2$ shift in direct space w.r.t. the CDW.

Considering now only the effect of the fourth order terms in $|\mathbf{\Psi}_0|$, if $\tilde{\beta}_2>0$, the free energy is minimized for $\theta = \pi/4$, such that both PDW components have the same magnitude and TRS is preserved.
On the other hand, if $\tilde{\beta}_2<0$, the free energy is minimized for $\theta= \{0, \pi/2\}$, corresponding to one of the PDW components being zero, breaking TRS.  
In the presence of both induction and fourth-order terms, in the direct vicinity of the phase transition the $\kappa$ term is dominant, leading to $\theta=\pi/4$.
However, as the magnitude $|\Psi_{\rm b}|$ (and concomitantly $|\vec{\Psi}_0|$) increases, a deviation of $\theta$ from $\pi/4$ becomes energetically favourable if  $\tilde{\beta}_2<0$ and $\tilde{\beta}_2<-\tilde{\alpha}^3/(8 \kappa^2 |\Psi_{\rm b}|^2 |\vec{\Delta}_0|^2)$.
To demonstrate this numerically, we minimize the free energy $F_\Psi+F_{\Delta \Psi}$ for zero magnetic field and with a CDW  background $|\vec{\Delta}_0|$ taken to be constant in temperature for $T/T_{\rm c, \Psi} \leq 1$.
We choose parameters satisfying the condition above 
at a finite $T$, see Fig.~\ref{fig:Intertwining}(b).
Here, the two components $|\Psi_{\pm Q_0}|$ onset with equal magnitudes at $T_{\rm c,\Psi}$, such that TRS is preserved, however they deviate at a slightly lower temperature, leading to TRSB on the surface.
Notice that the sign of the deviation, or equivalently the sign of $\theta-\pi/4$, is field trainable via the $\rho$ term in Eq.~(\ref{eq:FreeEnergyCoupled}), potentially explaining the field-trainable polar Kerr effect reported in Wei et al.~\cite{Wei2022}.

\emph{Imbalanced Fourier peaks.} Finally, we address the observations related to the Fourier peaks of the CDW around the points $\pm L_{1,2}$ above the superconducting ordering temperature.
To this end, we invoke ``higher harmonics" of the CDW order parameter with momenta related by RLVs, $\pm \mathbf{Q}_{1,2}=\mp \mathbf{Q}_0\pm \vec{q}_{1,2}$.
As $M_y \mathbf{Q}_{1,2}=-\mathbf{Q}_{2,1}$,  we organize the corresponding new order parameter harmonics into ``left" and ``right" groups, reminiscent of their location in reciprocal space, see Fig.~\ref{fig:SurfaceSymmetry} (b).
We define $\mathbf{\Delta}_{\rm L}=(\Delta_{+Q_1},\Delta_{-Q_2})$ and $\mathbf{\Delta}_{\rm R}=(\Delta_{+Q_2},\Delta_{-Q_1})$, with the same symmetry properties as $\mathbf{\Delta}_0$.
Note that these two-component order parameters naturally mix components at momenta $\vec{Q}_{1,2}$; what is at the core of the strong correlation between amplitude maps corresponding to Fourier peaks in the vicinity of $L_{1,2}$ reported in Lafleur et al.~\cite{LaFleur2024}.

\begin{figure}[t]
    \centering
    \includegraphics[width=0.98\linewidth]{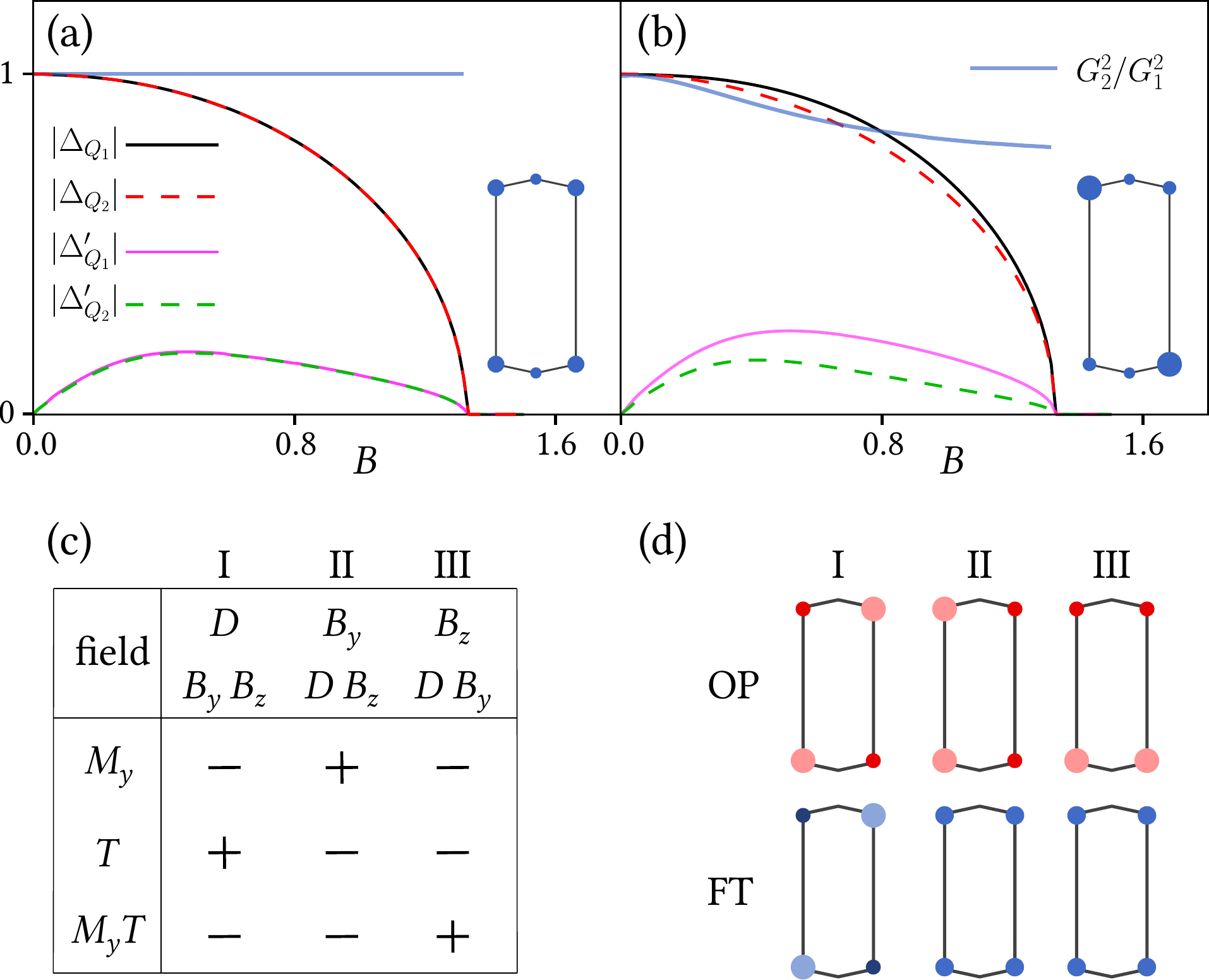}
    \caption{
    (a)-(b) Magnetic field evolution of charge and current order parameter amplitudes, and asymmetry of  Fourier peaks, captured by the ratio $G_2^2/G_1^2$ (schematic in inset). We used
    $-\alpha_{LR}/|\alpha|=\alpha_{LR}'/|\alpha|=\beta_{LR}/\beta=\zeta/\beta=\zeta'/\beta=1$, $\eta_\pm^{(\prime)}/|\alpha|=0.5$, $\nu_{1,2}/\sqrt{|\alpha|\beta}=0.2$, and normalized the order parameters with their zero-field amplitudes.
    In (a) and (b) $(\cos \chi \cdot \lambda_{\rm L,R}) /\sqrt{|\alpha|\beta}=0$ and 0.1, respectively. 
    (c) Summary of the symmetry properties of external symmetry breaking fields and their bilinears.
    (d) Schematic representation of the origin of the Fourier-peak imbalance or lack thereof for each case identified in (c).
    }
    \label{fig:F12Asymmetry}
\end{figure}

Accounting for these higher harmonics of the CDW, we can address the dependence of the corresponding Fourier peak amplitudes in the presence of magnetic field, which were reported to be suppressed at unequal rates as magnetic field is increased~\cite{Aishwarya2023,Aishwarya2024}. 
As the STM measurements probe a (real) amplitude modulation, the Fourier decomposition yields equal weights for $\pm \mathbf{Q}_i$, allowing us to focus on the peaks with $\mathbf{Q}_{1,2}$.
It would be natural to attribute the imbalance in the Fourier peaks to $\vec{B}_{\parallel M} \neq 0$, as this field is odd under $M_y$.
Unintuitively, the gap amplitudes  $G^2_{1,2}$ remain symmetric, such that $G_1^2/G_2^2=1$, in presence of a homogeneous $\vec{B}_{\parallel M}$,
see Fig.~\ref{fig:F12Asymmetry} (a).
This can be understood by the fact that the momenta $\mathbf{Q}_{1,2}$ are connected by the composition of TRS and $M_y$, and $\vec{B}_{\parallel M}$ being odd under both operations preserves their product.

To address the origin of Fourier peaks imbalance, we write the free energy for the higher harmonics of the order parameters $\Delta_{\rm L,R}^{(\prime)}$ as $F_{\rm LR}=F_{\rm LR}^0 + F_{\rm LR}^{B_z} +F_{\rm LR}^{B_y}$, with
\begin{align}
    F_{\rm LR}^0&= \sum_{i={\rm L,R}}\left(\frac{\alpha_{\rm LR}}{2} |\mathbf{\Delta}_i|^2  + \frac{\alpha_{\rm LR}'}{2} |\mathbf{\Delta}'_i|^2  + \frac{\beta_{\rm LR}}{4} |\mathbf{\Delta}_i|^4 \right)   \\
    &+ \Big( \eta_{\rm LR}\mathbf{\Delta}_{\rm R}^\dag \mathbf{\Delta}_{\rm L}  - 
    \eta_{\rm LR}'\mathbf{\Delta}_{\rm R}^{\prime \dag} \mathbf{\Delta}'_{\rm L} + {\rm h.c.} \Big) \nonumber,
\end{align}
where $\eta_{\rm LR}^{(\prime)}=(\eta_+^{(\prime)} -i\eta_-^{(\prime)})/2$ and we take $\alpha_{\rm LR}<0$, assuming $T<T_{c,\Delta}$ and $\eta^{(\prime)}_{LR}>0$. We only include the simplest fourth-order term for stability, with $\beta_{\rm LR}>0$. As before, $\alpha_{\rm LR}'>0$, such that currents are only induced in the presence of a magnetic field.
The evolution of $\Delta_{\rm L,R}$ for finite $B_z$ is captured by $F_{\rm LR}^{B_z}=F_{\rm L}^{B_z}+F_{\rm R}^{B_z}$ with
\begin{align}
    F_{i={\rm L,R}}^{B_z}=& B_z^2 \bigg( \frac{\zeta_{\rm LR}}{2} |\mathbf{\Delta}_i|^2+
    \frac{\zeta'_{\rm LR}}{2}|\mathbf{\Delta}'_i|^2 \bigg)  
    +B_z \nu_{i} \mathbf{\Delta}_{i}^\dagger \sigma_3 \mathbf{\Delta}_{i}',
\end{align}
where $\nu_{\rm L,R}=\nu_1\pm i \nu_2$, and we assume $\zeta_{\rm LR}^{(\prime)}$,  $\nu_{1,2}>0$.
We minimize $F_{\rm LR}$ for a range of $B_z$, see Fig.~\ref{fig:F12Asymmetry}(a), and find $|\Delta_{Q_1}|=|\Delta_{Q_2}|$ and $|\Delta_{Q_1}'|=|\Delta_{Q_2}'|$, which, therefore, does not account for the Fourier peak asymmetry.

As discussed above, the Fouries-peak asymmetry must be a consequence of extra symmetry breaking fields beyond $B_z$, which is odd under both $M_y$ and TRS. 
We summarize the distinct symmetry-breaking fields in Fig.~\ref{fig:F12Asymmetry}(c)-(d), where $B_z$ is indicated as type III.
In contrast, type I breaks only $M_y$ while preserving TRS, and can be associated with spatial inhomogeneities, accounted here by the external field $D$.
Finally, type II breaks only TRS and preserves $M_y$, associated with a magnetic-field component $B_y = B \cos\chi \neq 0$ ($\chi$ is the angle between the magnetic field and the mirror plane).
For the asymmetry between peaks at $\vec{Q}_{1,2}$ to manifest, the sum of the order parameter amplitudes at $\pm \vec{Q}_1$ and those at $\pm \vec{Q}_2$ have to be unequal.
This can be achieved by either a single  type-I field, or the simultaneous presence of a type-II and a type-III field.

As an illustrative example, we focus on the case with the magnetic field tilted away from the mirror plane, where the field component $B_y$ enters the free energy via
\begin{align}
F_{\rm LR}^{B_y}=B_y \Big[\lambda_{\rm L} \mathbf{\Delta}_{\rm L}^\dag \mathbf{\Delta}_L' - \lambda_R \mathbf{\Delta}_{\rm R}^\dag \mathbf{\Delta}_{\rm R}' \Big],
\end{align}
where $\lambda_{\rm L,R}=\lambda_1 \pm i \lambda_2$.
Minimizing $F_{\rm LR}$ with $B_y,B_z\neq 0$, we find a field-induced splitting of $|\Delta_{\pm Q_{1,2}}|$, as well as a pronounced deviation of $G_2^2/G_1^2$ from one as a function of field, see Fig.~\ref{fig:F12Asymmetry} (b).

\emph{Discussion and outlook.} 
In this letter, we employed a symmetry-based phenomenological analysis to account for salient experimental observations on the surface of UTe$_2$. We now summarize our results and propose targeted experiments to validate our proposal of intertwined surface and bulk orders. First, we proposed an attractive coupling between the surface CDW and bulk superconducting orders as the origin of the apparent suppression of the CDW at the superconducting upper critical field.
Such coupling between two phases positively renormalizes the nominal critical field and temperature of the subdominant order parameter. 
If the CDW is restricted to the surface, we expect superconductivity on the surface to be more robust than in the bulk.
Comparing $T_{\rm c, \Psi}$ on the surface with that in the bulk could therefore be a test for this scenario.

We also propose a novel mechanism for TRSB superconductivity, based on the multicomponent nature of the induced PDW order parameter on the surface.
If the presence of the CDW is surface dependent, our conjecture can be tested by cleaving the sample along a different plane, devoid of charge order.
According to our proposal, such a surface would not give rise to TRSB superconductivity. This could be verified by polar Kerr effect.

At last, our symmetry analysis shows that the asymmetric field evolution of the Fourier peaks at $\vec{Q}_{1,2}$ cannot be simply attributed to the presence of magnetic field perpendicular to the surface. Rather, it relies on further symmetry breaking effects, or a tilt of the magnetic field out of the mirror plane.
This scenario could be tested, for example, by sweeping the angle $\chi$, upon which the asymmetry should be minimal for $\chi=0$.
Furthermore, we also expect weaker asymmetry if the Fourier transform is performed on a field of view with less inhomogeneities.

Altogether, our analysis provides a comprehensive account of the symmetries associated with order parameters emerging on the surface of UTe$_2$ and puts forward a scenario with intertwined orders emerging on the surface and in the bulk. The experimental tests we propose to prove this scenario could further our understanding of the surface physics of UTe$_2$, and have the potential to shed light on the outstanding puzzle surrounding the nature and origin of the TRSB in this fascinating compound.

%%%%%%%%%%%%%%%%%%%%%%%%%%%%%%%%%%%%%%%
%%%%%%%%%%%%%%%%%%%%%%%%%%%%%%%%%%%%%%%

\begin{acknowledgments}
A.S. is grateful for financial support from the Swiss National Science Foundation (SNSF) through Division II (No. 184739). 
A.R. also acknowledges financial support from the SNSF through an Ambizione Grant No. 186043.
\end{acknowledgments}

%%%%%%%%%%%%%%%%%%%%%%%%%%%%%%%%%%%%%%%%%
%%%%%%%%%%%%%%%%%%%%%%%%%%%%%%%%%%%%%%%%%
%%%%%%%%%%%%%%%%%%%%%%%%%%%%%%%%%%%%%%%%%
\bibliography{bibliography}

\begin{thebibliography}{38}%
\makeatletter
\providecommand \@ifxundefined [1]{%
 \@ifx{#1\undefined}
}%
\providecommand \@ifnum [1]{%
 \ifnum #1\expandafter \@firstoftwo
 \else \expandafter \@secondoftwo
 \fi
}%
\providecommand \@ifx [1]{%
 \ifx #1\expandafter \@firstoftwo
 \else \expandafter \@secondoftwo
 \fi
}%
\providecommand \natexlab [1]{#1}%
\providecommand \enquote  [1]{``#1''}%
\providecommand \bibnamefont  [1]{#1}%
\providecommand \bibfnamefont [1]{#1}%
\providecommand \citenamefont [1]{#1}%
\providecommand \href@noop [0]{\@secondoftwo}%
\providecommand \href [0]{\begingroup \@sanitize@url \@href}%
\providecommand \@href[1]{\@@startlink{#1}\@@href}%
\providecommand \@@href[1]{\endgroup#1\@@endlink}%
\providecommand \@sanitize@url [0]{\catcode `\\12\catcode `\$12\catcode
  `\&12\catcode `\#12\catcode `\^12\catcode `\_12\catcode `\%12\relax}%
\providecommand \@@startlink[1]{}%
\providecommand \@@endlink[0]{}%
\providecommand \url  [0]{\begingroup\@sanitize@url \@url }%
\providecommand \@url [1]{\endgroup\@href {#1}{\urlprefix }}%
\providecommand \urlprefix  [0]{URL }%
\providecommand \Eprint [0]{\href }%
\providecommand \doibase [0]{https://doi.org/}%
\providecommand \selectlanguage [0]{\@gobble}%
\providecommand \bibinfo  [0]{\@secondoftwo}%
\providecommand \bibfield  [0]{\@secondoftwo}%
\providecommand \translation [1]{[#1]}%
\providecommand \BibitemOpen [0]{}%
\providecommand \bibitemStop [0]{}%
\providecommand \bibitemNoStop [0]{.\EOS\space}%
\providecommand \EOS [0]{\spacefactor3000\relax}%
\providecommand \BibitemShut  [1]{\csname bibitem#1\endcsname}%
\let\auto@bib@innerbib\@empty
%</preamble>
\bibitem [{\citenamefont {Fradkin}\ \emph {et~al.}(2015)\citenamefont
  {Fradkin}, \citenamefont {Kivelson},\ and\ \citenamefont
  {Tranquada}}]{Fradkin2015}%
  \BibitemOpen
  \bibfield  {author} {\bibinfo {author} {\bibfnamefont {E.}~\bibnamefont
  {Fradkin}}, \bibinfo {author} {\bibfnamefont {S.~A.}\ \bibnamefont
  {Kivelson}},\ and\ \bibinfo {author} {\bibfnamefont {J.~M.}\ \bibnamefont
  {Tranquada}},\ }\href {https://doi.org/10.1103/RevModPhys.87.457} {\bibfield
  {journal} {\bibinfo  {journal} {Rev. Mod. Phys.}\ }\textbf {\bibinfo {volume}
  {87}},\ \bibinfo {pages} {457} (\bibinfo {year} {2015})}\BibitemShut
  {NoStop}%
\bibitem [{\citenamefont {Fernandes}\ \emph {et~al.}(2019)\citenamefont
  {Fernandes}, \citenamefont {Orth},\ and\ \citenamefont
  {Schmalian}}]{Fernandes2019}%
  \BibitemOpen
  \bibfield  {author} {\bibinfo {author} {\bibfnamefont {R.~M.}\ \bibnamefont
  {Fernandes}}, \bibinfo {author} {\bibfnamefont {P.~P.}\ \bibnamefont
  {Orth}},\ and\ \bibinfo {author} {\bibfnamefont {J.}~\bibnamefont
  {Schmalian}},\ }\href
  {https://doi.org/10.1146/annurev-conmatphys-031218-013200} {\bibfield
  {journal} {\bibinfo  {journal} {Annual Review of Condensed Matter Physics}\
  }\textbf {\bibinfo {volume} {10}},\ \bibinfo {pages} {133–154} (\bibinfo
  {year} {2019})}\BibitemShut {NoStop}%
\bibitem [{\citenamefont {Agterberg}\ \emph {et~al.}(2020)\citenamefont
  {Agterberg}, \citenamefont {Davis}, \citenamefont {Edkins}, \citenamefont
  {Fradkin}, \citenamefont {Van~Harlingen}, \citenamefont {Kivelson},
  \citenamefont {Lee}, \citenamefont {Radzihovsky}, \citenamefont {Tranquada},\
  and\ \citenamefont {Wang}}]{Agterberg2020}%
  \BibitemOpen
  \bibfield  {author} {\bibinfo {author} {\bibfnamefont {D.~F.}\ \bibnamefont
  {Agterberg}}, \bibinfo {author} {\bibfnamefont {J.~S.}\ \bibnamefont
  {Davis}}, \bibinfo {author} {\bibfnamefont {S.~D.}\ \bibnamefont {Edkins}},
  \bibinfo {author} {\bibfnamefont {E.}~\bibnamefont {Fradkin}}, \bibinfo
  {author} {\bibfnamefont {D.~J.}\ \bibnamefont {Van~Harlingen}}, \bibinfo
  {author} {\bibfnamefont {S.~A.}\ \bibnamefont {Kivelson}}, \bibinfo {author}
  {\bibfnamefont {P.~A.}\ \bibnamefont {Lee}}, \bibinfo {author} {\bibfnamefont
  {L.}~\bibnamefont {Radzihovsky}}, \bibinfo {author} {\bibfnamefont {J.~M.}\
  \bibnamefont {Tranquada}},\ and\ \bibinfo {author} {\bibfnamefont
  {Y.}~\bibnamefont {Wang}},\ }\href
  {https://doi.org/10.1146/annurev-conmatphys-031119-050711} {\bibfield
  {journal} {\bibinfo  {journal} {Annual Review of Condensed Matter Physics}\
  }\textbf {\bibinfo {volume} {11}},\ \bibinfo {pages} {231–270} (\bibinfo
  {year} {2020})}\BibitemShut {NoStop}%
\bibitem [{\citenamefont {Agterberg}\ and\ \citenamefont
  {Tsunetsugu}(2008)}]{Agterberg2008}%
  \BibitemOpen
  \bibfield  {author} {\bibinfo {author} {\bibfnamefont {D.~F.}\ \bibnamefont
  {Agterberg}}\ and\ \bibinfo {author} {\bibfnamefont {H.}~\bibnamefont
  {Tsunetsugu}},\ }\href {https://doi.org/10.1038/nphys999} {\bibfield
  {journal} {\bibinfo  {journal} {Nature Physics}\ }\textbf {\bibinfo {volume}
  {4}},\ \bibinfo {pages} {639–642} (\bibinfo {year} {2008})}\BibitemShut
  {NoStop}%
\bibitem [{\citenamefont {Berg}\ \emph
  {et~al.}(2009{\natexlab{a}})\citenamefont {Berg}, \citenamefont {Fradkin},\
  and\ \citenamefont {Kivelson}}]{Berg2009NP}%
  \BibitemOpen
  \bibfield  {author} {\bibinfo {author} {\bibfnamefont {E.}~\bibnamefont
  {Berg}}, \bibinfo {author} {\bibfnamefont {E.}~\bibnamefont {Fradkin}},\ and\
  \bibinfo {author} {\bibfnamefont {S.~A.}\ \bibnamefont {Kivelson}},\ }\href
  {https://doi.org/10.1038/nphys1389} {\bibfield  {journal} {\bibinfo
  {journal} {Nature Physics}\ }\textbf {\bibinfo {volume} {5}},\ \bibinfo
  {pages} {830–833} (\bibinfo {year} {2009}{\natexlab{a}})}\BibitemShut
  {NoStop}%
\bibitem [{\citenamefont {Berg}\ \emph
  {et~al.}(2009{\natexlab{b}})\citenamefont {Berg}, \citenamefont {Fradkin},
  \citenamefont {Kivelson},\ and\ \citenamefont {Tranquada}}]{Berg2009NJP}%
  \BibitemOpen
  \bibfield  {author} {\bibinfo {author} {\bibfnamefont {E.}~\bibnamefont
  {Berg}}, \bibinfo {author} {\bibfnamefont {E.}~\bibnamefont {Fradkin}},
  \bibinfo {author} {\bibfnamefont {S.~A.}\ \bibnamefont {Kivelson}},\ and\
  \bibinfo {author} {\bibfnamefont {J.~M.}\ \bibnamefont {Tranquada}},\ }\href
  {https://doi.org/10.1088/1367-2630/11/11/115004} {\bibfield  {journal}
  {\bibinfo  {journal} {New Journal of Physics}\ }\textbf {\bibinfo {volume}
  {11}},\ \bibinfo {pages} {115004} (\bibinfo {year}
  {2009}{\natexlab{b}})}\BibitemShut {NoStop}%
\bibitem [{\citenamefont {Hamidian}\ \emph {et~al.}(2016)\citenamefont
  {Hamidian}, \citenamefont {Edkins}, \citenamefont {Joo}, \citenamefont
  {Kostin}, \citenamefont {Eisaki}, \citenamefont {Uchida}, \citenamefont
  {Lawler}, \citenamefont {Kim}, \citenamefont {Mackenzie}, \citenamefont
  {Fujita}, \citenamefont {Lee},\ and\ \citenamefont {Davis}}]{Hamidian2016}%
  \BibitemOpen
  \bibfield  {author} {\bibinfo {author} {\bibfnamefont {M.~H.}\ \bibnamefont
  {Hamidian}}, \bibinfo {author} {\bibfnamefont {S.~D.}\ \bibnamefont
  {Edkins}}, \bibinfo {author} {\bibfnamefont {S.~H.}\ \bibnamefont {Joo}},
  \bibinfo {author} {\bibfnamefont {A.}~\bibnamefont {Kostin}}, \bibinfo
  {author} {\bibfnamefont {H.}~\bibnamefont {Eisaki}}, \bibinfo {author}
  {\bibfnamefont {S.}~\bibnamefont {Uchida}}, \bibinfo {author} {\bibfnamefont
  {M.~J.}\ \bibnamefont {Lawler}}, \bibinfo {author} {\bibfnamefont {E.-A.}\
  \bibnamefont {Kim}}, \bibinfo {author} {\bibfnamefont {A.~P.}\ \bibnamefont
  {Mackenzie}}, \bibinfo {author} {\bibfnamefont {K.}~\bibnamefont {Fujita}},
  \bibinfo {author} {\bibfnamefont {J.}~\bibnamefont {Lee}},\ and\ \bibinfo
  {author} {\bibfnamefont {J.~C.~S.}\ \bibnamefont {Davis}},\ }\href
  {https://doi.org/10.1038/nature17411} {\bibfield  {journal} {\bibinfo
  {journal} {Nature}\ }\textbf {\bibinfo {volume} {532}},\ \bibinfo {pages}
  {343–347} (\bibinfo {year} {2016})}\BibitemShut {NoStop}%
\bibitem [{\citenamefont {Ruan}\ \emph {et~al.}(2018)\citenamefont {Ruan},
  \citenamefont {Li}, \citenamefont {Hu}, \citenamefont {Hao}, \citenamefont
  {Li}, \citenamefont {Cai}, \citenamefont {Zhou}, \citenamefont {Lee},\ and\
  \citenamefont {Wang}}]{Ruan2018}%
  \BibitemOpen
  \bibfield  {author} {\bibinfo {author} {\bibfnamefont {W.}~\bibnamefont
  {Ruan}}, \bibinfo {author} {\bibfnamefont {X.}~\bibnamefont {Li}}, \bibinfo
  {author} {\bibfnamefont {C.}~\bibnamefont {Hu}}, \bibinfo {author}
  {\bibfnamefont {Z.}~\bibnamefont {Hao}}, \bibinfo {author} {\bibfnamefont
  {H.}~\bibnamefont {Li}}, \bibinfo {author} {\bibfnamefont {P.}~\bibnamefont
  {Cai}}, \bibinfo {author} {\bibfnamefont {X.}~\bibnamefont {Zhou}}, \bibinfo
  {author} {\bibfnamefont {D.-H.}\ \bibnamefont {Lee}},\ and\ \bibinfo {author}
  {\bibfnamefont {Y.}~\bibnamefont {Wang}},\ }\href
  {https://doi.org/10.1038/s41567-018-0276-8} {\bibfield  {journal} {\bibinfo
  {journal} {Nature Physics}\ }\textbf {\bibinfo {volume} {14}},\ \bibinfo
  {pages} {1178–1182} (\bibinfo {year} {2018})}\BibitemShut {NoStop}%
\bibitem [{\citenamefont {Edkins}\ \emph {et~al.}(2019)\citenamefont {Edkins},
  \citenamefont {Kostin}, \citenamefont {Fujita}, \citenamefont {Mackenzie},
  \citenamefont {Eisaki}, \citenamefont {Uchida}, \citenamefont {Sachdev},
  \citenamefont {Lawler}, \citenamefont {Kim}, \citenamefont {Séamus~Davis},\
  and\ \citenamefont {Hamidian}}]{Edkins2019}%
  \BibitemOpen
  \bibfield  {author} {\bibinfo {author} {\bibfnamefont {S.~D.}\ \bibnamefont
  {Edkins}}, \bibinfo {author} {\bibfnamefont {A.}~\bibnamefont {Kostin}},
  \bibinfo {author} {\bibfnamefont {K.}~\bibnamefont {Fujita}}, \bibinfo
  {author} {\bibfnamefont {A.~P.}\ \bibnamefont {Mackenzie}}, \bibinfo {author}
  {\bibfnamefont {H.}~\bibnamefont {Eisaki}}, \bibinfo {author} {\bibfnamefont
  {S.}~\bibnamefont {Uchida}}, \bibinfo {author} {\bibfnamefont
  {S.}~\bibnamefont {Sachdev}}, \bibinfo {author} {\bibfnamefont {M.~J.}\
  \bibnamefont {Lawler}}, \bibinfo {author} {\bibfnamefont {E.-A.}\
  \bibnamefont {Kim}}, \bibinfo {author} {\bibfnamefont {J.~C.}\ \bibnamefont
  {Séamus~Davis}},\ and\ \bibinfo {author} {\bibfnamefont {M.~H.}\
  \bibnamefont {Hamidian}},\ }\href {https://doi.org/10.1126/science.aat1773}
  {\bibfield  {journal} {\bibinfo  {journal} {Science}\ }\textbf {\bibinfo
  {volume} {364}},\ \bibinfo {pages} {976–980} (\bibinfo {year}
  {2019})}\BibitemShut {NoStop}%
\bibitem [{\citenamefont {Liu}\ \emph {et~al.}(2021)\citenamefont {Liu},
  \citenamefont {Chong}, \citenamefont {Sharma},\ and\ \citenamefont
  {Davis}}]{Liu2021}%
  \BibitemOpen
  \bibfield  {author} {\bibinfo {author} {\bibfnamefont {X.}~\bibnamefont
  {Liu}}, \bibinfo {author} {\bibfnamefont {Y.~X.}\ \bibnamefont {Chong}},
  \bibinfo {author} {\bibfnamefont {R.}~\bibnamefont {Sharma}},\ and\ \bibinfo
  {author} {\bibfnamefont {J.~C.~S.}\ \bibnamefont {Davis}},\ }\href
  {https://doi.org/10.1126/science.abd4607} {\bibfield  {journal} {\bibinfo
  {journal} {Science}\ }\textbf {\bibinfo {volume} {372}},\ \bibinfo {pages}
  {1447–1452} (\bibinfo {year} {2021})}\BibitemShut {NoStop}%
\bibitem [{\citenamefont {Chen}\ \emph {et~al.}(2021)\citenamefont {Chen},
  \citenamefont {Yang}, \citenamefont {Hu}, \citenamefont {Zhao}, \citenamefont
  {Yuan}, \citenamefont {Xing}, \citenamefont {Qian}, \citenamefont {Huang},
  \citenamefont {Li}, \citenamefont {Ye}, \citenamefont {Ma}, \citenamefont
  {Ni}, \citenamefont {Zhang}, \citenamefont {Yin}, \citenamefont {Gong},
  \citenamefont {Tu}, \citenamefont {Lei}, \citenamefont {Tan}, \citenamefont
  {Zhou}, \citenamefont {Shen}, \citenamefont {Dong}, \citenamefont {Yan},
  \citenamefont {Wang},\ and\ \citenamefont {Gao}}]{Chen2021}%
  \BibitemOpen
  \bibfield  {author} {\bibinfo {author} {\bibfnamefont {H.}~\bibnamefont
  {Chen}}, \bibinfo {author} {\bibfnamefont {H.}~\bibnamefont {Yang}}, \bibinfo
  {author} {\bibfnamefont {B.}~\bibnamefont {Hu}}, \bibinfo {author}
  {\bibfnamefont {Z.}~\bibnamefont {Zhao}}, \bibinfo {author} {\bibfnamefont
  {J.}~\bibnamefont {Yuan}}, \bibinfo {author} {\bibfnamefont {Y.}~\bibnamefont
  {Xing}}, \bibinfo {author} {\bibfnamefont {G.}~\bibnamefont {Qian}}, \bibinfo
  {author} {\bibfnamefont {Z.}~\bibnamefont {Huang}}, \bibinfo {author}
  {\bibfnamefont {G.}~\bibnamefont {Li}}, \bibinfo {author} {\bibfnamefont
  {Y.}~\bibnamefont {Ye}}, \bibinfo {author} {\bibfnamefont {S.}~\bibnamefont
  {Ma}}, \bibinfo {author} {\bibfnamefont {S.}~\bibnamefont {Ni}}, \bibinfo
  {author} {\bibfnamefont {H.}~\bibnamefont {Zhang}}, \bibinfo {author}
  {\bibfnamefont {Q.}~\bibnamefont {Yin}}, \bibinfo {author} {\bibfnamefont
  {C.}~\bibnamefont {Gong}}, \bibinfo {author} {\bibfnamefont {Z.}~\bibnamefont
  {Tu}}, \bibinfo {author} {\bibfnamefont {H.}~\bibnamefont {Lei}}, \bibinfo
  {author} {\bibfnamefont {H.}~\bibnamefont {Tan}}, \bibinfo {author}
  {\bibfnamefont {S.}~\bibnamefont {Zhou}}, \bibinfo {author} {\bibfnamefont
  {C.}~\bibnamefont {Shen}}, \bibinfo {author} {\bibfnamefont {X.}~\bibnamefont
  {Dong}}, \bibinfo {author} {\bibfnamefont {B.}~\bibnamefont {Yan}}, \bibinfo
  {author} {\bibfnamefont {Z.}~\bibnamefont {Wang}},\ and\ \bibinfo {author}
  {\bibfnamefont {H.-J.}\ \bibnamefont {Gao}},\ }\href
  {https://doi.org/10.1038/s41586-021-03983-5} {\bibfield  {journal} {\bibinfo
  {journal} {Nature}\ }\textbf {\bibinfo {volume} {599}},\ \bibinfo {pages}
  {222–228} (\bibinfo {year} {2021})}\BibitemShut {NoStop}%
\bibitem [{\citenamefont {Gu}\ \emph {et~al.}(2023)\citenamefont {Gu},
  \citenamefont {Carroll}, \citenamefont {Wang}, \citenamefont {Ran},
  \citenamefont {Broyles}, \citenamefont {Siddiquee}, \citenamefont {Butch},
  \citenamefont {Saha}, \citenamefont {Paglione}, \citenamefont {Davis},\ and\
  \citenamefont {Liu}}]{Gu2023}%
  \BibitemOpen
  \bibfield  {author} {\bibinfo {author} {\bibfnamefont {Q.}~\bibnamefont
  {Gu}}, \bibinfo {author} {\bibfnamefont {J.~P.}\ \bibnamefont {Carroll}},
  \bibinfo {author} {\bibfnamefont {S.}~\bibnamefont {Wang}}, \bibinfo {author}
  {\bibfnamefont {S.}~\bibnamefont {Ran}}, \bibinfo {author} {\bibfnamefont
  {C.}~\bibnamefont {Broyles}}, \bibinfo {author} {\bibfnamefont
  {H.}~\bibnamefont {Siddiquee}}, \bibinfo {author} {\bibfnamefont {N.~P.}\
  \bibnamefont {Butch}}, \bibinfo {author} {\bibfnamefont {S.~R.}\ \bibnamefont
  {Saha}}, \bibinfo {author} {\bibfnamefont {J.}~\bibnamefont {Paglione}},
  \bibinfo {author} {\bibfnamefont {J.~C.~S.}\ \bibnamefont {Davis}},\ and\
  \bibinfo {author} {\bibfnamefont {X.}~\bibnamefont {Liu}},\ }\href
  {https://doi.org/10.1038/s41586-023-05919-7} {\bibfield  {journal} {\bibinfo
  {journal} {Nature}\ }\textbf {\bibinfo {volume} {618}},\ \bibinfo {pages}
  {921–927} (\bibinfo {year} {2023})}\BibitemShut {NoStop}%
\bibitem [{\citenamefont {Aishwarya}\ \emph {et~al.}(2023)\citenamefont
  {Aishwarya}, \citenamefont {May-Mann}, \citenamefont {Raghavan},
  \citenamefont {Nie}, \citenamefont {Romanelli}, \citenamefont {Ran},
  \citenamefont {Saha}, \citenamefont {Paglione}, \citenamefont {Butch},
  \citenamefont {Fradkin},\ and\ \citenamefont {Madhavan}}]{Aishwarya2023}%
  \BibitemOpen
  \bibfield  {author} {\bibinfo {author} {\bibfnamefont {A.}~\bibnamefont
  {Aishwarya}}, \bibinfo {author} {\bibfnamefont {J.}~\bibnamefont {May-Mann}},
  \bibinfo {author} {\bibfnamefont {A.}~\bibnamefont {Raghavan}}, \bibinfo
  {author} {\bibfnamefont {L.}~\bibnamefont {Nie}}, \bibinfo {author}
  {\bibfnamefont {M.}~\bibnamefont {Romanelli}}, \bibinfo {author}
  {\bibfnamefont {S.}~\bibnamefont {Ran}}, \bibinfo {author} {\bibfnamefont
  {S.~R.}\ \bibnamefont {Saha}}, \bibinfo {author} {\bibfnamefont
  {J.}~\bibnamefont {Paglione}}, \bibinfo {author} {\bibfnamefont {N.~P.}\
  \bibnamefont {Butch}}, \bibinfo {author} {\bibfnamefont {E.}~\bibnamefont
  {Fradkin}},\ and\ \bibinfo {author} {\bibfnamefont {V.}~\bibnamefont
  {Madhavan}},\ }\href {https://doi.org/10.1038/s41586-023-06005-8} {\bibfield
  {journal} {\bibinfo  {journal} {Nature}\ }\textbf {\bibinfo {volume} {618}},\
  \bibinfo {pages} {928–933} (\bibinfo {year} {2023})}\BibitemShut {NoStop}%
\bibitem [{\citenamefont {LaFleur}\ \emph {et~al.}(2024)\citenamefont
  {LaFleur}, \citenamefont {Li}, \citenamefont {Frank}, \citenamefont {Xu},
  \citenamefont {Cheng}, \citenamefont {Wang}, \citenamefont {Butch},\ and\
  \citenamefont {Zeljkovic}}]{LaFleur2024}%
  \BibitemOpen
  \bibfield  {author} {\bibinfo {author} {\bibfnamefont {A.}~\bibnamefont
  {LaFleur}}, \bibinfo {author} {\bibfnamefont {H.}~\bibnamefont {Li}},
  \bibinfo {author} {\bibfnamefont {C.~E.}\ \bibnamefont {Frank}}, \bibinfo
  {author} {\bibfnamefont {M.}~\bibnamefont {Xu}}, \bibinfo {author}
  {\bibfnamefont {S.}~\bibnamefont {Cheng}}, \bibinfo {author} {\bibfnamefont
  {Z.}~\bibnamefont {Wang}}, \bibinfo {author} {\bibfnamefont {N.~P.}\
  \bibnamefont {Butch}},\ and\ \bibinfo {author} {\bibfnamefont
  {I.}~\bibnamefont {Zeljkovic}},\ }\bibfield  {journal} {\bibinfo  {journal}
  {Nature Communications}\ }\textbf {\bibinfo {volume} {15}},\ \href
  {https://doi.org/10.1038/s41467-024-48844-7} {10.1038/s41467-024-48844-7}
  (\bibinfo {year} {2024})\BibitemShut {NoStop}%
\bibitem [{\citenamefont {Kengle}\ \emph
  {et~al.}(2024{\natexlab{a}})\citenamefont {Kengle}, \citenamefont {Vonka},
  \citenamefont {Francoual}, \citenamefont {Chang}, \citenamefont {Abbamonte},
  \citenamefont {Janoschek}, \citenamefont {Rosa},\ and\ \citenamefont
  {Simeth}}]{Kengle2024natcom}%
  \BibitemOpen
  \bibfield  {author} {\bibinfo {author} {\bibfnamefont {C.~S.}\ \bibnamefont
  {Kengle}}, \bibinfo {author} {\bibfnamefont {J.}~\bibnamefont {Vonka}},
  \bibinfo {author} {\bibfnamefont {S.}~\bibnamefont {Francoual}}, \bibinfo
  {author} {\bibfnamefont {J.}~\bibnamefont {Chang}}, \bibinfo {author}
  {\bibfnamefont {P.}~\bibnamefont {Abbamonte}}, \bibinfo {author}
  {\bibfnamefont {M.}~\bibnamefont {Janoschek}}, \bibinfo {author}
  {\bibfnamefont {P.}~\bibnamefont {Rosa}},\ and\ \bibinfo {author}
  {\bibfnamefont {W.}~\bibnamefont {Simeth}},\ }\href@noop {} {\bibfield
  {journal} {\bibinfo  {journal} {Nature communications}\ }\textbf {\bibinfo
  {volume} {15}},\ \bibinfo {pages} {9713} (\bibinfo {year}
  {2024}{\natexlab{a}})}\BibitemShut {NoStop}%
\bibitem [{\citenamefont {Theuss}\ \emph
  {et~al.}(2024{\natexlab{a}})\citenamefont {Theuss}, \citenamefont {Shragai},
  \citenamefont {Grissonnanche}, \citenamefont {Peralta}, \citenamefont
  {Simarro}, \citenamefont {Hayes}, \citenamefont {Saha}, \citenamefont {Eo},
  \citenamefont {Suarez}, \citenamefont {Salinas}, \citenamefont {Pokharel},
  \citenamefont {Wilson}, \citenamefont {Butch}, \citenamefont {Paglione},\
  and\ \citenamefont {Ramshaw}}]{Theuss2024}%
  \BibitemOpen
  \bibfield  {author} {\bibinfo {author} {\bibfnamefont {F.}~\bibnamefont
  {Theuss}}, \bibinfo {author} {\bibfnamefont {A.}~\bibnamefont {Shragai}},
  \bibinfo {author} {\bibfnamefont {G.}~\bibnamefont {Grissonnanche}}, \bibinfo
  {author} {\bibfnamefont {L.}~\bibnamefont {Peralta}}, \bibinfo {author}
  {\bibfnamefont {G.~d. l.~F.}\ \bibnamefont {Simarro}}, \bibinfo {author}
  {\bibfnamefont {I.~M.}\ \bibnamefont {Hayes}}, \bibinfo {author}
  {\bibfnamefont {S.~R.}\ \bibnamefont {Saha}}, \bibinfo {author}
  {\bibfnamefont {Y.~S.}\ \bibnamefont {Eo}}, \bibinfo {author} {\bibfnamefont
  {A.}~\bibnamefont {Suarez}}, \bibinfo {author} {\bibfnamefont {A.~C.}\
  \bibnamefont {Salinas}}, \bibinfo {author} {\bibfnamefont {G.}~\bibnamefont
  {Pokharel}}, \bibinfo {author} {\bibfnamefont {S.~D.}\ \bibnamefont
  {Wilson}}, \bibinfo {author} {\bibfnamefont {N.~P.}\ \bibnamefont {Butch}},
  \bibinfo {author} {\bibfnamefont {J.}~\bibnamefont {Paglione}},\ and\
  \bibinfo {author} {\bibfnamefont {B.~J.}\ \bibnamefont {Ramshaw}},\ }\href
  {https://doi.org/10.1103/PhysRevB.110.144507} {\bibfield  {journal} {\bibinfo
   {journal} {Phys. Rev. B}\ }\textbf {\bibinfo {volume} {110}},\ \bibinfo
  {pages} {144507} (\bibinfo {year} {2024}{\natexlab{a}})}\BibitemShut
  {NoStop}%
\bibitem [{\citenamefont {Kengle}\ \emph
  {et~al.}(2024{\natexlab{b}})\citenamefont {Kengle}, \citenamefont
  {Chaudhuri}, \citenamefont {Guo}, \citenamefont {Johnson}, \citenamefont
  {Bettler}, \citenamefont {Simeth}, \citenamefont {Krogstad}, \citenamefont
  {Islam}, \citenamefont {Ran}, \citenamefont {Saha}, \citenamefont {Paglione},
  \citenamefont {Butch}, \citenamefont {Fradkin}, \citenamefont {Madhavan},\
  and\ \citenamefont {Abbamonte}}]{Kengle2024}%
  \BibitemOpen
  \bibfield  {author} {\bibinfo {author} {\bibfnamefont {C.~S.}\ \bibnamefont
  {Kengle}}, \bibinfo {author} {\bibfnamefont {D.}~\bibnamefont {Chaudhuri}},
  \bibinfo {author} {\bibfnamefont {X.}~\bibnamefont {Guo}}, \bibinfo {author}
  {\bibfnamefont {T.~A.}\ \bibnamefont {Johnson}}, \bibinfo {author}
  {\bibfnamefont {S.}~\bibnamefont {Bettler}}, \bibinfo {author} {\bibfnamefont
  {W.}~\bibnamefont {Simeth}}, \bibinfo {author} {\bibfnamefont {M.~J.}\
  \bibnamefont {Krogstad}}, \bibinfo {author} {\bibfnamefont {Z.}~\bibnamefont
  {Islam}}, \bibinfo {author} {\bibfnamefont {S.}~\bibnamefont {Ran}}, \bibinfo
  {author} {\bibfnamefont {S.~R.}\ \bibnamefont {Saha}}, \bibinfo {author}
  {\bibfnamefont {J.}~\bibnamefont {Paglione}}, \bibinfo {author}
  {\bibfnamefont {N.~P.}\ \bibnamefont {Butch}}, \bibinfo {author}
  {\bibfnamefont {E.}~\bibnamefont {Fradkin}}, \bibinfo {author} {\bibfnamefont
  {V.}~\bibnamefont {Madhavan}},\ and\ \bibinfo {author} {\bibfnamefont
  {P.}~\bibnamefont {Abbamonte}},\ }\href
  {https://doi.org/10.1103/PhysRevB.110.145101} {\bibfield  {journal} {\bibinfo
   {journal} {Phys. Rev. B}\ }\textbf {\bibinfo {volume} {110}},\ \bibinfo
  {pages} {145101} (\bibinfo {year} {2024}{\natexlab{b}})}\BibitemShut
  {NoStop}%
\bibitem [{\citenamefont {Ran}\ \emph {et~al.}(2019)\citenamefont {Ran},
  \citenamefont {Eckberg}, \citenamefont {Ding}, \citenamefont {Furukawa},
  \citenamefont {Metz}, \citenamefont {Saha}, \citenamefont {Liu},
  \citenamefont {Zic}, \citenamefont {Kim}, \citenamefont {Paglione},\ and\
  \citenamefont {Butch}}]{Ran2019}%
  \BibitemOpen
  \bibfield  {author} {\bibinfo {author} {\bibfnamefont {S.}~\bibnamefont
  {Ran}}, \bibinfo {author} {\bibfnamefont {C.}~\bibnamefont {Eckberg}},
  \bibinfo {author} {\bibfnamefont {Q.-P.}\ \bibnamefont {Ding}}, \bibinfo
  {author} {\bibfnamefont {Y.}~\bibnamefont {Furukawa}}, \bibinfo {author}
  {\bibfnamefont {T.}~\bibnamefont {Metz}}, \bibinfo {author} {\bibfnamefont
  {S.~R.}\ \bibnamefont {Saha}}, \bibinfo {author} {\bibfnamefont {I.-L.}\
  \bibnamefont {Liu}}, \bibinfo {author} {\bibfnamefont {M.}~\bibnamefont
  {Zic}}, \bibinfo {author} {\bibfnamefont {H.}~\bibnamefont {Kim}}, \bibinfo
  {author} {\bibfnamefont {J.}~\bibnamefont {Paglione}},\ and\ \bibinfo
  {author} {\bibfnamefont {N.~P.}\ \bibnamefont {Butch}},\ }\href
  {https://doi.org/10.1126/science.aav8645} {\bibfield  {journal} {\bibinfo
  {journal} {Science}\ }\textbf {\bibinfo {volume} {365}},\ \bibinfo {pages}
  {684–687} (\bibinfo {year} {2019})}\BibitemShut {NoStop}%
\bibitem [{\citenamefont {Aoki}\ \emph {et~al.}(2019)\citenamefont {Aoki},
  \citenamefont {Nakamura}, \citenamefont {Honda}, \citenamefont {Li},
  \citenamefont {Homma}, \citenamefont {Shimizu}, \citenamefont {Sato},
  \citenamefont {Knebel}, \citenamefont {Brison}, \citenamefont {Pourret},
  \citenamefont {Braithwaite}, \citenamefont {Lapertot}, \citenamefont {Niu},
  \citenamefont {Vališka}, \citenamefont {Harima},\ and\ \citenamefont
  {Flouquet}}]{Aoki2019}%
  \BibitemOpen
  \bibfield  {author} {\bibinfo {author} {\bibfnamefont {D.}~\bibnamefont
  {Aoki}}, \bibinfo {author} {\bibfnamefont {A.}~\bibnamefont {Nakamura}},
  \bibinfo {author} {\bibfnamefont {F.}~\bibnamefont {Honda}}, \bibinfo
  {author} {\bibfnamefont {D.}~\bibnamefont {Li}}, \bibinfo {author}
  {\bibfnamefont {Y.}~\bibnamefont {Homma}}, \bibinfo {author} {\bibfnamefont
  {Y.}~\bibnamefont {Shimizu}}, \bibinfo {author} {\bibfnamefont {Y.~J.}\
  \bibnamefont {Sato}}, \bibinfo {author} {\bibfnamefont {G.}~\bibnamefont
  {Knebel}}, \bibinfo {author} {\bibfnamefont {J.-P.}\ \bibnamefont {Brison}},
  \bibinfo {author} {\bibfnamefont {A.}~\bibnamefont {Pourret}}, \bibinfo
  {author} {\bibfnamefont {D.}~\bibnamefont {Braithwaite}}, \bibinfo {author}
  {\bibfnamefont {G.}~\bibnamefont {Lapertot}}, \bibinfo {author}
  {\bibfnamefont {Q.}~\bibnamefont {Niu}}, \bibinfo {author} {\bibfnamefont
  {M.}~\bibnamefont {Vališka}}, \bibinfo {author} {\bibfnamefont
  {H.}~\bibnamefont {Harima}},\ and\ \bibinfo {author} {\bibfnamefont
  {J.}~\bibnamefont {Flouquet}},\ }\href
  {https://doi.org/10.7566/jpsj.88.043702} {\bibfield  {journal} {\bibinfo
  {journal} {Journal of the Physical Society of Japan}\ }\textbf {\bibinfo
  {volume} {88}},\ \bibinfo {pages} {043702} (\bibinfo {year}
  {2019})}\BibitemShut {NoStop}%
\bibitem [{\citenamefont {Nakamine}\ \emph {et~al.}(2019)\citenamefont
  {Nakamine}, \citenamefont {Kitagawa}, \citenamefont {Ishida}, \citenamefont
  {Tokunaga}, \citenamefont {Sakai}, \citenamefont {Kambe}, \citenamefont
  {Nakamura}, \citenamefont {Shimizu}, \citenamefont {Homma}, \citenamefont
  {Li}, \citenamefont {Honda},\ and\ \citenamefont {Aoki}}]{Nakamine2019}%
  \BibitemOpen
  \bibfield  {author} {\bibinfo {author} {\bibfnamefont {G.}~\bibnamefont
  {Nakamine}}, \bibinfo {author} {\bibfnamefont {S.}~\bibnamefont {Kitagawa}},
  \bibinfo {author} {\bibfnamefont {K.}~\bibnamefont {Ishida}}, \bibinfo
  {author} {\bibfnamefont {Y.}~\bibnamefont {Tokunaga}}, \bibinfo {author}
  {\bibfnamefont {H.}~\bibnamefont {Sakai}}, \bibinfo {author} {\bibfnamefont
  {S.}~\bibnamefont {Kambe}}, \bibinfo {author} {\bibfnamefont
  {A.}~\bibnamefont {Nakamura}}, \bibinfo {author} {\bibfnamefont
  {Y.}~\bibnamefont {Shimizu}}, \bibinfo {author} {\bibfnamefont
  {Y.}~\bibnamefont {Homma}}, \bibinfo {author} {\bibfnamefont
  {D.}~\bibnamefont {Li}}, \bibinfo {author} {\bibfnamefont {F.}~\bibnamefont
  {Honda}},\ and\ \bibinfo {author} {\bibfnamefont {D.}~\bibnamefont {Aoki}},\
  }\href {https://doi.org/10.7566/jpsj.88.113703} {\bibfield  {journal}
  {\bibinfo  {journal} {Journal of the Physical Society of Japan}\ }\textbf
  {\bibinfo {volume} {88}},\ \bibinfo {pages} {113703} (\bibinfo {year}
  {2019})}\BibitemShut {NoStop}%
\bibitem [{\citenamefont {Sundar}\ \emph {et~al.}(2019)\citenamefont {Sundar},
  \citenamefont {Gheidi}, \citenamefont {Akintola}, \citenamefont {C\^ot\'e},
  \citenamefont {Dunsiger}, \citenamefont {Ran}, \citenamefont {Butch},
  \citenamefont {Saha}, \citenamefont {Paglione},\ and\ \citenamefont
  {Sonier}}]{Sundar2019}%
  \BibitemOpen
  \bibfield  {author} {\bibinfo {author} {\bibfnamefont {S.}~\bibnamefont
  {Sundar}}, \bibinfo {author} {\bibfnamefont {S.}~\bibnamefont {Gheidi}},
  \bibinfo {author} {\bibfnamefont {K.}~\bibnamefont {Akintola}}, \bibinfo
  {author} {\bibfnamefont {A.~M.}\ \bibnamefont {C\^ot\'e}}, \bibinfo {author}
  {\bibfnamefont {S.~R.}\ \bibnamefont {Dunsiger}}, \bibinfo {author}
  {\bibfnamefont {S.}~\bibnamefont {Ran}}, \bibinfo {author} {\bibfnamefont
  {N.~P.}\ \bibnamefont {Butch}}, \bibinfo {author} {\bibfnamefont {S.~R.}\
  \bibnamefont {Saha}}, \bibinfo {author} {\bibfnamefont {J.}~\bibnamefont
  {Paglione}},\ and\ \bibinfo {author} {\bibfnamefont {J.~E.}\ \bibnamefont
  {Sonier}},\ }\href {https://doi.org/10.1103/PhysRevB.100.140502} {\bibfield
  {journal} {\bibinfo  {journal} {Phys. Rev. B}\ }\textbf {\bibinfo {volume}
  {100}},\ \bibinfo {pages} {140502} (\bibinfo {year} {2019})}\BibitemShut
  {NoStop}%
\bibitem [{\citenamefont {Azari}\ \emph {et~al.}(2023)\citenamefont {Azari},
  \citenamefont {Yakovlev}, \citenamefont {Rye}, \citenamefont {Dunsiger},
  \citenamefont {Sundar}, \citenamefont {Bordelon}, \citenamefont {Thomas},
  \citenamefont {Thompson}, \citenamefont {Rosa},\ and\ \citenamefont
  {Sonier}}]{Azari2023}%
  \BibitemOpen
  \bibfield  {author} {\bibinfo {author} {\bibfnamefont {N.}~\bibnamefont
  {Azari}}, \bibinfo {author} {\bibfnamefont {M.}~\bibnamefont {Yakovlev}},
  \bibinfo {author} {\bibfnamefont {N.}~\bibnamefont {Rye}}, \bibinfo {author}
  {\bibfnamefont {S.~R.}\ \bibnamefont {Dunsiger}}, \bibinfo {author}
  {\bibfnamefont {S.}~\bibnamefont {Sundar}}, \bibinfo {author} {\bibfnamefont
  {M.~M.}\ \bibnamefont {Bordelon}}, \bibinfo {author} {\bibfnamefont {S.~M.}\
  \bibnamefont {Thomas}}, \bibinfo {author} {\bibfnamefont {J.~D.}\
  \bibnamefont {Thompson}}, \bibinfo {author} {\bibfnamefont {P.~F.~S.}\
  \bibnamefont {Rosa}},\ and\ \bibinfo {author} {\bibfnamefont {J.~E.}\
  \bibnamefont {Sonier}},\ }\href
  {https://doi.org/10.1103/PhysRevLett.131.226504} {\bibfield  {journal}
  {\bibinfo  {journal} {Phys. Rev. Lett.}\ }\textbf {\bibinfo {volume} {131}},\
  \bibinfo {pages} {226504} (\bibinfo {year} {2023})}\BibitemShut {NoStop}%
\bibitem [{\citenamefont {Hayes}\ \emph {et~al.}(2021)\citenamefont {Hayes},
  \citenamefont {Wei}, \citenamefont {Metz}, \citenamefont {Zhang},
  \citenamefont {Eo}, \citenamefont {Ran}, \citenamefont {Saha}, \citenamefont
  {Collini}, \citenamefont {Butch}, \citenamefont {Agterberg}, \citenamefont
  {Kapitulnik},\ and\ \citenamefont {Paglione}}]{Hayes2021}%
  \BibitemOpen
  \bibfield  {author} {\bibinfo {author} {\bibfnamefont {I.~M.}\ \bibnamefont
  {Hayes}}, \bibinfo {author} {\bibfnamefont {D.~S.}\ \bibnamefont {Wei}},
  \bibinfo {author} {\bibfnamefont {T.}~\bibnamefont {Metz}}, \bibinfo {author}
  {\bibfnamefont {J.}~\bibnamefont {Zhang}}, \bibinfo {author} {\bibfnamefont
  {Y.~S.}\ \bibnamefont {Eo}}, \bibinfo {author} {\bibfnamefont
  {S.}~\bibnamefont {Ran}}, \bibinfo {author} {\bibfnamefont {S.~R.}\
  \bibnamefont {Saha}}, \bibinfo {author} {\bibfnamefont {J.}~\bibnamefont
  {Collini}}, \bibinfo {author} {\bibfnamefont {N.~P.}\ \bibnamefont {Butch}},
  \bibinfo {author} {\bibfnamefont {D.~F.}\ \bibnamefont {Agterberg}}, \bibinfo
  {author} {\bibfnamefont {A.}~\bibnamefont {Kapitulnik}},\ and\ \bibinfo
  {author} {\bibfnamefont {J.}~\bibnamefont {Paglione}},\ }\href
  {https://doi.org/10.1126/science.abb0272} {\bibfield  {journal} {\bibinfo
  {journal} {Science}\ }\textbf {\bibinfo {volume} {373}},\ \bibinfo {pages}
  {797–801} (\bibinfo {year} {2021})}\BibitemShut {NoStop}%
\bibitem [{\citenamefont {Wei}\ \emph {et~al.}(2022)\citenamefont {Wei},
  \citenamefont {Saykin}, \citenamefont {Miller}, \citenamefont {Ran},
  \citenamefont {Saha}, \citenamefont {Agterberg}, \citenamefont {Schmalian},
  \citenamefont {Butch}, \citenamefont {Paglione},\ and\ \citenamefont
  {Kapitulnik}}]{Wei2022}%
  \BibitemOpen
  \bibfield  {author} {\bibinfo {author} {\bibfnamefont {D.~S.}\ \bibnamefont
  {Wei}}, \bibinfo {author} {\bibfnamefont {D.}~\bibnamefont {Saykin}},
  \bibinfo {author} {\bibfnamefont {O.~Y.}\ \bibnamefont {Miller}}, \bibinfo
  {author} {\bibfnamefont {S.}~\bibnamefont {Ran}}, \bibinfo {author}
  {\bibfnamefont {S.~R.}\ \bibnamefont {Saha}}, \bibinfo {author}
  {\bibfnamefont {D.~F.}\ \bibnamefont {Agterberg}}, \bibinfo {author}
  {\bibfnamefont {J.}~\bibnamefont {Schmalian}}, \bibinfo {author}
  {\bibfnamefont {N.~P.}\ \bibnamefont {Butch}}, \bibinfo {author}
  {\bibfnamefont {J.}~\bibnamefont {Paglione}},\ and\ \bibinfo {author}
  {\bibfnamefont {A.}~\bibnamefont {Kapitulnik}},\ }\href
  {https://doi.org/10.1103/PhysRevB.105.024521} {\bibfield  {journal} {\bibinfo
   {journal} {Phys. Rev. B}\ }\textbf {\bibinfo {volume} {105}},\ \bibinfo
  {pages} {024521} (\bibinfo {year} {2022})}\BibitemShut {NoStop}%
\bibitem [{\citenamefont {Jiao}\ \emph {et~al.}(2020)\citenamefont {Jiao},
  \citenamefont {Howard}, \citenamefont {Ran}, \citenamefont {Wang},
  \citenamefont {Rodriguez}, \citenamefont {Sigrist}, \citenamefont {Wang},
  \citenamefont {Butch},\ and\ \citenamefont {Madhavan}}]{Jiao2020}%
  \BibitemOpen
  \bibfield  {author} {\bibinfo {author} {\bibfnamefont {L.}~\bibnamefont
  {Jiao}}, \bibinfo {author} {\bibfnamefont {S.}~\bibnamefont {Howard}},
  \bibinfo {author} {\bibfnamefont {S.}~\bibnamefont {Ran}}, \bibinfo {author}
  {\bibfnamefont {Z.}~\bibnamefont {Wang}}, \bibinfo {author} {\bibfnamefont
  {J.~O.}\ \bibnamefont {Rodriguez}}, \bibinfo {author} {\bibfnamefont
  {M.}~\bibnamefont {Sigrist}}, \bibinfo {author} {\bibfnamefont
  {Z.}~\bibnamefont {Wang}}, \bibinfo {author} {\bibfnamefont {N.~P.}\
  \bibnamefont {Butch}},\ and\ \bibinfo {author} {\bibfnamefont
  {V.}~\bibnamefont {Madhavan}},\ }\href
  {https://doi.org/10.1038/s41586-020-2122-2} {\bibfield  {journal} {\bibinfo
  {journal} {Nature}\ }\textbf {\bibinfo {volume} {579}},\ \bibinfo {pages}
  {523–527} (\bibinfo {year} {2020})}\BibitemShut {NoStop}%
\bibitem [{\citenamefont {Ishihara}\ \emph {et~al.}(2023)\citenamefont
  {Ishihara}, \citenamefont {Roppongi}, \citenamefont {Kobayashi},
  \citenamefont {Imamura}, \citenamefont {Mizukami}, \citenamefont {Sakai},
  \citenamefont {Opletal}, \citenamefont {Tokiwa}, \citenamefont {Haga},
  \citenamefont {Hashimoto},\ and\ \citenamefont {Shibauchi}}]{Ishihara2023}%
  \BibitemOpen
  \bibfield  {author} {\bibinfo {author} {\bibfnamefont {K.}~\bibnamefont
  {Ishihara}}, \bibinfo {author} {\bibfnamefont {M.}~\bibnamefont {Roppongi}},
  \bibinfo {author} {\bibfnamefont {M.}~\bibnamefont {Kobayashi}}, \bibinfo
  {author} {\bibfnamefont {K.}~\bibnamefont {Imamura}}, \bibinfo {author}
  {\bibfnamefont {Y.}~\bibnamefont {Mizukami}}, \bibinfo {author}
  {\bibfnamefont {H.}~\bibnamefont {Sakai}}, \bibinfo {author} {\bibfnamefont
  {P.}~\bibnamefont {Opletal}}, \bibinfo {author} {\bibfnamefont
  {Y.}~\bibnamefont {Tokiwa}}, \bibinfo {author} {\bibfnamefont
  {Y.}~\bibnamefont {Haga}}, \bibinfo {author} {\bibfnamefont {K.}~\bibnamefont
  {Hashimoto}},\ and\ \bibinfo {author} {\bibfnamefont {T.}~\bibnamefont
  {Shibauchi}},\ }\bibfield  {journal} {\bibinfo  {journal} {Nature
  Communications}\ }\textbf {\bibinfo {volume} {14}},\ \href
  {https://doi.org/10.1038/s41467-023-38688-y} {10.1038/s41467-023-38688-y}
  (\bibinfo {year} {2023})\BibitemShut {NoStop}%
\bibitem [{\citenamefont {Aoki}\ \emph {et~al.}(2022)\citenamefont {Aoki},
  \citenamefont {Brison}, \citenamefont {Flouquet}, \citenamefont {Ishida},
  \citenamefont {Knebel}, \citenamefont {Tokunaga},\ and\ \citenamefont
  {Yanase}}]{Aoki2022}%
  \BibitemOpen
  \bibfield  {author} {\bibinfo {author} {\bibfnamefont {D.}~\bibnamefont
  {Aoki}}, \bibinfo {author} {\bibfnamefont {J.-P.}\ \bibnamefont {Brison}},
  \bibinfo {author} {\bibfnamefont {J.}~\bibnamefont {Flouquet}}, \bibinfo
  {author} {\bibfnamefont {K.}~\bibnamefont {Ishida}}, \bibinfo {author}
  {\bibfnamefont {G.}~\bibnamefont {Knebel}}, \bibinfo {author} {\bibfnamefont
  {Y.}~\bibnamefont {Tokunaga}},\ and\ \bibinfo {author} {\bibfnamefont
  {Y.}~\bibnamefont {Yanase}},\ }\href
  {https://doi.org/10.1088/1361-648x/ac5863} {\bibfield  {journal} {\bibinfo
  {journal} {Journal of Physics: Condensed Matter}\ }\textbf {\bibinfo {volume}
  {34}},\ \bibinfo {pages} {243002} (\bibinfo {year} {2022})}\BibitemShut
  {NoStop}%
\bibitem [{\citenamefont {Theuss}\ \emph
  {et~al.}(2024{\natexlab{b}})\citenamefont {Theuss}, \citenamefont {Shragai},
  \citenamefont {Grissonnanche}, \citenamefont {Hayes}, \citenamefont {Saha},
  \citenamefont {Eo}, \citenamefont {Suarez}, \citenamefont {Shishidou},
  \citenamefont {Butch}, \citenamefont {Paglione},\ and\ \citenamefont
  {Ramshaw}}]{Theuss2024NaturePhysics}%
  \BibitemOpen
  \bibfield  {author} {\bibinfo {author} {\bibfnamefont {F.}~\bibnamefont
  {Theuss}}, \bibinfo {author} {\bibfnamefont {A.}~\bibnamefont {Shragai}},
  \bibinfo {author} {\bibfnamefont {G.}~\bibnamefont {Grissonnanche}}, \bibinfo
  {author} {\bibfnamefont {I.~M.}\ \bibnamefont {Hayes}}, \bibinfo {author}
  {\bibfnamefont {S.~R.}\ \bibnamefont {Saha}}, \bibinfo {author}
  {\bibfnamefont {Y.~S.}\ \bibnamefont {Eo}}, \bibinfo {author} {\bibfnamefont
  {A.}~\bibnamefont {Suarez}}, \bibinfo {author} {\bibfnamefont
  {T.}~\bibnamefont {Shishidou}}, \bibinfo {author} {\bibfnamefont {N.~P.}\
  \bibnamefont {Butch}}, \bibinfo {author} {\bibfnamefont {J.}~\bibnamefont
  {Paglione}},\ and\ \bibinfo {author} {\bibfnamefont {B.~J.}\ \bibnamefont
  {Ramshaw}},\ }\href {https://doi.org/10.1038/s41567-024-02493-1} {\bibfield
  {journal} {\bibinfo  {journal} {Nature Physics}\ }\textbf {\bibinfo {volume}
  {20}},\ \bibinfo {pages} {1124–1130} (\bibinfo {year}
  {2024}{\natexlab{b}})}\BibitemShut {NoStop}%
\bibitem [{\citenamefont {Wang}\ \emph {et~al.}(2025)\citenamefont {Wang},
  \citenamefont {Zhussupbekov}, \citenamefont {Carroll}, \citenamefont {Hu},
  \citenamefont {Liu}, \citenamefont {Pangburn}, \citenamefont {Crepieux},
  \citenamefont {Pepin}, \citenamefont {Broyles}, \citenamefont {Ran},
  \citenamefont {Butch}, \citenamefont {Saha}, \citenamefont {Paglione},
  \citenamefont {Bena}, \citenamefont {Davis},\ and\ \citenamefont
  {Gu}}]{Wang2025}%
  \BibitemOpen
  \bibfield  {author} {\bibinfo {author} {\bibfnamefont {S.}~\bibnamefont
  {Wang}}, \bibinfo {author} {\bibfnamefont {K.}~\bibnamefont {Zhussupbekov}},
  \bibinfo {author} {\bibfnamefont {J.~P.}\ \bibnamefont {Carroll}}, \bibinfo
  {author} {\bibfnamefont {B.}~\bibnamefont {Hu}}, \bibinfo {author}
  {\bibfnamefont {X.}~\bibnamefont {Liu}}, \bibinfo {author} {\bibfnamefont
  {E.}~\bibnamefont {Pangburn}}, \bibinfo {author} {\bibfnamefont
  {A.}~\bibnamefont {Crepieux}}, \bibinfo {author} {\bibfnamefont
  {C.}~\bibnamefont {Pepin}}, \bibinfo {author} {\bibfnamefont
  {C.}~\bibnamefont {Broyles}}, \bibinfo {author} {\bibfnamefont
  {S.}~\bibnamefont {Ran}}, \bibinfo {author} {\bibfnamefont {N.~P.}\
  \bibnamefont {Butch}}, \bibinfo {author} {\bibfnamefont {S.}~\bibnamefont
  {Saha}}, \bibinfo {author} {\bibfnamefont {J.}~\bibnamefont {Paglione}},
  \bibinfo {author} {\bibfnamefont {C.}~\bibnamefont {Bena}}, \bibinfo {author}
  {\bibfnamefont {J.~C.~S.}\ \bibnamefont {Davis}},\ and\ \bibinfo {author}
  {\bibfnamefont {Q.}~\bibnamefont {Gu}},\ }\href
  {https://arxiv.org/abs/2503.17761} {\bibfield  {journal} {\bibinfo  {journal}
  {arXiv:2503.17761}\ } (\bibinfo {year} {2025})}\BibitemShut {NoStop}%
\bibitem [{\citenamefont {Gu}\ \emph {et~al.}(2025)\citenamefont {Gu},
  \citenamefont {Wang}, \citenamefont {Carroll}, \citenamefont {Zhussupbekov},
  \citenamefont {Broyles}, \citenamefont {Ran}, \citenamefont {Butch},
  \citenamefont {Saha}, \citenamefont {Paglione}, \citenamefont {Liu},
  \citenamefont {Davis},\ and\ \citenamefont {Lee}}]{Gu2025}%
  \BibitemOpen
  \bibfield  {author} {\bibinfo {author} {\bibfnamefont {Q.}~\bibnamefont
  {Gu}}, \bibinfo {author} {\bibfnamefont {S.}~\bibnamefont {Wang}}, \bibinfo
  {author} {\bibfnamefont {J.~P.}\ \bibnamefont {Carroll}}, \bibinfo {author}
  {\bibfnamefont {K.}~\bibnamefont {Zhussupbekov}}, \bibinfo {author}
  {\bibfnamefont {C.}~\bibnamefont {Broyles}}, \bibinfo {author} {\bibfnamefont
  {S.}~\bibnamefont {Ran}}, \bibinfo {author} {\bibfnamefont {N.~P.}\
  \bibnamefont {Butch}}, \bibinfo {author} {\bibfnamefont {S.}~\bibnamefont
  {Saha}}, \bibinfo {author} {\bibfnamefont {J.}~\bibnamefont {Paglione}},
  \bibinfo {author} {\bibfnamefont {X.}~\bibnamefont {Liu}}, \bibinfo {author}
  {\bibfnamefont {J.~C.~S.}\ \bibnamefont {Davis}},\ and\ \bibinfo {author}
  {\bibfnamefont {D.-H.}\ \bibnamefont {Lee}},\ }\href
  {https://arxiv.org/abs/2501.16636} {\bibfield  {journal} {\bibinfo  {journal}
  {arXiv:2501.16636}\ } (\bibinfo {year} {2025})}\BibitemShut {NoStop}%
\bibitem [{\citenamefont {Crépieux}\ \emph {et~al.}(2025)\citenamefont
  {Crépieux}, \citenamefont {Pangburn}, \citenamefont {Wang}, \citenamefont
  {Zhussupbekov}, \citenamefont {Carroll}, \citenamefont {Hu}, \citenamefont
  {Gu}, \citenamefont {Davis}, \citenamefont {Pépin},\ and\ \citenamefont
  {Bena}}]{Crépieux2025}%
  \BibitemOpen
  \bibfield  {author} {\bibinfo {author} {\bibfnamefont {A.}~\bibnamefont
  {Crépieux}}, \bibinfo {author} {\bibfnamefont {E.}~\bibnamefont {Pangburn}},
  \bibinfo {author} {\bibfnamefont {S.}~\bibnamefont {Wang}}, \bibinfo {author}
  {\bibfnamefont {K.}~\bibnamefont {Zhussupbekov}}, \bibinfo {author}
  {\bibfnamefont {J.~P.}\ \bibnamefont {Carroll}}, \bibinfo {author}
  {\bibfnamefont {B.}~\bibnamefont {Hu}}, \bibinfo {author} {\bibfnamefont
  {Q.}~\bibnamefont {Gu}}, \bibinfo {author} {\bibfnamefont {J.~C.~S.}\
  \bibnamefont {Davis}}, \bibinfo {author} {\bibfnamefont {C.}~\bibnamefont
  {Pépin}},\ and\ \bibinfo {author} {\bibfnamefont {C.}~\bibnamefont {Bena}},\
  }\href {https://arxiv.org/abs/2503.17762} {\bibfield  {journal} {\bibinfo
  {journal} {arXiv:2503.17762}\ } (\bibinfo {year} {2025})}\BibitemShut
  {NoStop}%
\bibitem [{\citenamefont {Christiansen}\ \emph {et~al.}(2025)\citenamefont
  {Christiansen}, \citenamefont {Geier}, \citenamefont {Andersen},\ and\
  \citenamefont {Kreisel}}]{Christiansen2025}%
  \BibitemOpen
  \bibfield  {author} {\bibinfo {author} {\bibfnamefont {H.}~\bibnamefont
  {Christiansen}}, \bibinfo {author} {\bibfnamefont {M.}~\bibnamefont {Geier}},
  \bibinfo {author} {\bibfnamefont {B.~M.}\ \bibnamefont {Andersen}},\ and\
  \bibinfo {author} {\bibfnamefont {A.}~\bibnamefont {Kreisel}},\ }\href
  {https://arxiv.org/abs/2503.11603} {\bibfield  {journal} {\bibinfo  {journal}
  {arXiv:2503.11603}\ } (\bibinfo {year} {2025})}\BibitemShut {NoStop}%
\bibitem [{\citenamefont {Hutanu}\ \emph {et~al.}(2020)\citenamefont {Hutanu},
  \citenamefont {Deng}, \citenamefont {Ran}, \citenamefont {Fuhrman},
  \citenamefont {Thoma},\ and\ \citenamefont {Butch}}]{Hutanu2020}%
  \BibitemOpen
  \bibfield  {author} {\bibinfo {author} {\bibfnamefont {V.}~\bibnamefont
  {Hutanu}}, \bibinfo {author} {\bibfnamefont {H.}~\bibnamefont {Deng}},
  \bibinfo {author} {\bibfnamefont {S.}~\bibnamefont {Ran}}, \bibinfo {author}
  {\bibfnamefont {W.~T.}\ \bibnamefont {Fuhrman}}, \bibinfo {author}
  {\bibfnamefont {H.}~\bibnamefont {Thoma}},\ and\ \bibinfo {author}
  {\bibfnamefont {N.~P.}\ \bibnamefont {Butch}},\ }\href
  {https://doi.org/10.1107/s2052520619016950} {\bibfield  {journal} {\bibinfo
  {journal} {Acta Crystallographica Section B Structural Science, Crystal
  Engineering and Materials}\ }\textbf {\bibinfo {volume} {76}},\ \bibinfo
  {pages} {137–143} (\bibinfo {year} {2020})}\BibitemShut {NoStop}%
\bibitem [{\citenamefont {Setyawan}\ and\ \citenamefont
  {Curtarolo}(2010)}]{Setyawan2010}%
  \BibitemOpen
  \bibfield  {author} {\bibinfo {author} {\bibfnamefont {W.}~\bibnamefont
  {Setyawan}}\ and\ \bibinfo {author} {\bibfnamefont {S.}~\bibnamefont
  {Curtarolo}},\ }\href {https://doi.org/10.1016/j.commatsci.2010.05.010}
  {\bibfield  {journal} {\bibinfo  {journal} {Computational Materials Science}\
  }\textbf {\bibinfo {volume} {49}},\ \bibinfo {pages} {299–312} (\bibinfo
  {year} {2010})}\BibitemShut {NoStop}%
\bibitem [{\citenamefont {Sigrist}\ and\ \citenamefont
  {Ueda}(1991)}]{Sigrist1991}%
  \BibitemOpen
  \bibfield  {author} {\bibinfo {author} {\bibfnamefont {M.}~\bibnamefont
  {Sigrist}}\ and\ \bibinfo {author} {\bibfnamefont {K.}~\bibnamefont {Ueda}},\
  }\href {https://doi.org/10.1103/RevModPhys.63.239} {\bibfield  {journal}
  {\bibinfo  {journal} {Rev. Mod. Phys.}\ }\textbf {\bibinfo {volume} {63}},\
  \bibinfo {pages} {239} (\bibinfo {year} {1991})}\BibitemShut {NoStop}%
\bibitem [{\citenamefont {Kallin}\ and\ \citenamefont
  {Berlinsky}(2016)}]{Kallin2016}%
  \BibitemOpen
  \bibfield  {author} {\bibinfo {author} {\bibfnamefont {C.}~\bibnamefont
  {Kallin}}\ and\ \bibinfo {author} {\bibfnamefont {J.}~\bibnamefont
  {Berlinsky}},\ }\href {https://doi.org/10.1088/0034-4885/79/5/054502}
  {\bibfield  {journal} {\bibinfo  {journal} {Reports on Progress in Physics}\
  }\textbf {\bibinfo {volume} {79}},\ \bibinfo {pages} {054502} (\bibinfo
  {year} {2016})}\BibitemShut {NoStop}%
\bibitem [{\citenamefont {Ramires}(2022)}]{Ramires2022}%
  \BibitemOpen
  \bibfield  {author} {\bibinfo {author} {\bibfnamefont {A.}~\bibnamefont
  {Ramires}},\ }\href {https://doi.org/10.1080/00107514.2022.2140499}
  {\bibfield  {journal} {\bibinfo  {journal} {Contemporary Physics}\ }\textbf
  {\bibinfo {volume} {63}},\ \bibinfo {pages} {71–86} (\bibinfo {year}
  {2022})}\BibitemShut {NoStop}%
\bibitem [{\citenamefont {Aishwarya}\ \emph {et~al.}(2024)\citenamefont
  {Aishwarya}, \citenamefont {May-Mann}, \citenamefont {Almoalem},
  \citenamefont {Ran}, \citenamefont {Saha}, \citenamefont {Paglione},
  \citenamefont {Butch}, \citenamefont {Fradkin},\ and\ \citenamefont
  {Madhavan}}]{Aishwarya2024}%
  \BibitemOpen
  \bibfield  {author} {\bibinfo {author} {\bibfnamefont {A.}~\bibnamefont
  {Aishwarya}}, \bibinfo {author} {\bibfnamefont {J.}~\bibnamefont {May-Mann}},
  \bibinfo {author} {\bibfnamefont {A.}~\bibnamefont {Almoalem}}, \bibinfo
  {author} {\bibfnamefont {S.}~\bibnamefont {Ran}}, \bibinfo {author}
  {\bibfnamefont {S.~R.}\ \bibnamefont {Saha}}, \bibinfo {author}
  {\bibfnamefont {J.}~\bibnamefont {Paglione}}, \bibinfo {author}
  {\bibfnamefont {N.~P.}\ \bibnamefont {Butch}}, \bibinfo {author}
  {\bibfnamefont {E.}~\bibnamefont {Fradkin}},\ and\ \bibinfo {author}
  {\bibfnamefont {V.}~\bibnamefont {Madhavan}},\ }\href
  {https://doi.org/10.1038/s41567-024-02429-9} {\bibfield  {journal} {\bibinfo
  {journal} {Nature Physics}\ }\textbf {\bibinfo {volume} {20}},\ \bibinfo
  {pages} {964} (\bibinfo {year} {2024})}\BibitemShut {NoStop}%
\end{thebibliography}%

\end{document}